\documentclass[lettersize,journal]{IEEEtran}
\usepackage{amsmath}
\usepackage{graphicx}
\usepackage{amsfonts}
\usepackage{caption}
\usepackage{color}
\usepackage{cuted}
\usepackage{multicol}
\usepackage{algorithmic}
\usepackage{algorithm}
\usepackage{float}
\usepackage{comment}
\usepackage{amssymb}
\usepackage{stfloats}
\usepackage{bbm}
\usepackage[caption=false]{subfig}
\usepackage{cleveref}
\usepackage[english]{babel}
\usepackage{amsthm}
\usepackage{bigints}

\newcommand{\tx}{\mathrm{tx}}

\theoremstyle{plain}
\newtheorem{thm}{Theorem}

\theoremstyle{definition}

\theoremstyle{remark}

\usepackage{pifont}
\newcommand{\cmark}{\text{\ding{51}}}
\newcommand{\xmark}{\text{\ding{55}}}

\graphicspath{{Figures/}{../Figures/}}

\begin{document}

\title{Closed-Form Expressions for I/O Relation in Zak-OTFS with Different Delay-Doppler Filters}

\author{Arpan Das,~Fathima Jesbin,~and Ananthanarayanan Chockalingam \\
Department of ECE, Indian Institute of Science, Bangalore 560012 
}

\maketitle

\begin{abstract}
The transceiver operations in the delay-Doppler (DD) domain in Zak-OTFS modulation, including DD domain filtering at the transmitter and receiver, involve twisted convolution operation. The twisted convolution operations give rise to multiple integrals in the end-to-end DD domain input-output (I/O) relation. The I/O relation plays a crucial role in performance evaluation and algorithm development for transceiver implementation. In this paper, we derive discrete DD domain closed-form expressions for the I/O relation and noise covariance in Zak-OTFS. We derive these expressions for sinc and Gaussian pulse shaping DD filters at the transmitter (Tx). On the receiver (Rx) side, three types of DD filters are considered, viz., $(i)$ Rx filter identical to Tx filter (referred to as `identical filtering'), $(ii)$ Rx filter matched to the Tx filter (referred to as `matched filtering'), and $(iii)$ Rx filter matched to both Tx filter and channel response (referred to as `channel matched filtering'). For all the above cases, except for the case of sinc identical filtering, we derive exact I/O relation and noise covariance expressions in closed-form. For the sinc identical filtering case, we derive approximate closed-form expressions which are shown to be accurate. Using the derived closed-form expressions, we evaluate the bit error performance of Zak-OTFS for different Tx/Rx filter configurations. Our results using Vehicular-A (Veh-A) channel model with fractional DDs show that, while matched filtering achieves slightly better or almost same performance as identical filtering, channel matched filtering achieves the best performance among the three.
\end{abstract}


\section{Introduction}
\label{sec1}
\IEEEPARstart{N}
EXT generation mobile communication systems (e.g., 6G) strive to ensure reliable communication in high-mobility scenarios, while catering to the emerging radar sensing needs \cite{ref1},\cite{ref2a}. High-mobility scenarios result in channel characteristics that are rapidly time-varying \cite{bello}, rendering the channels as time-selective (due to Doppler spread) in addition to being frequency-selective (due to delay spread). Dealing with high time-selectivity in wireless channels is challenging. Orthogonal time frequency space (OTFS) waveform is a promising waveform that can efficiently serve the communication (through information multiplexing in the delay-Doppler domain) as well as the radar sensing needs in 6G and beyond \cite{otfs_hadani1}-\cite{isac_otfs}. OTFS is known to offer significantly superior performance compared to OFDM, which is prone to inter-carrier interference caused by time-selectivity of the channel. 

OTFS research so far has evolved in two phases. In the early version of OTFS introduced in 2017 \cite{otfs_hadani1}, the delay-Doppler (DD) domain to time domain conversion (and vice versa) is done in two steps, namely, DD domain to time-frequency (TF) domain conversion using inverse symplectic finite Fourier transform followed by TF domain to time domain conversion using Heisenberg transform. This approach was motivated by compatibility with existing multicarrier (MC) modulation in 4G/5G (hence it is referred to as ``MC-OTFS'') \cite{otfs_hadani1}-\cite{zak_r4}. Recently, a more fundamental approach of direct conversion from DD domain to time domain using a Zak theoretic framework (referred to as ``Zak-OTFS'') has emerged \cite{ref3},\cite{ref4}. Two key aspects are central to Zak-OTFS. First, it provides a formal mathematical framework using Zak theory for describing OTFS and studying its fundamental properties, in a manner analogous to how Fourier theory constitutes an appropriate mathematical framework for describing and understanding OFDM. Second, Zak-OTFS waveform is more robust to a larger range of delay and Doppler spreads of the channel. This is because the input-output (I/O) relation in Zak-OTFS is non-fading and predictable, even in the presence of significant delay and Doppler spreads, and, as a consequence, the channel can be efficiently acquired and equalized.

While MC-OTFS research is  more mature, Zak-OTFS research is relatively more recent and new \cite{ref3}-\cite{zak_otfs_book}. In this context, we note that the early derivation of  a closed-form expression for the end-to-end I/O relation in the DD domain for MC-OTFS in \cite{mp_det} triggered a surge of MC-OTFS research output, exploiting the I/O relation expression for performance analysis \cite{analysis1}, signal detection and channel estimation \cite{mimo_otfs}-\cite{h_b_mishra}, precoding \cite{precoding1}, adoption to spatial modulation \cite{zak_r3} and space-time shift keying \cite{zak_r4}, etc., in single user and multiuser environments. A key and new contribution in this paper is the derivation of such closed-form expressions in compact form for the end-to-end I/O relation in Zak-OTFS, which has not been reported. The context, relevance, and usefulness of the contributions are highlighted below.

The transceiver operations in the DD domain in Zak-OTFS modulation, including DD domain filtering at the transmitter and receiver, involve twisted convolution operation\footnote{Twisted convolution of two DD functions $a(\tau,\nu)$ and $b(\tau,\nu)$, where $\tau$, $\nu$ denote the delay and Doppler variables, respectively, is defined as \\
$a(\tau,\nu) \ast_\sigma b(\tau,\nu) \overset{\Delta}{=} \iint a(\tau', \nu') b(\tau-\tau',\nu-\nu')e^{j2\pi\nu'(\tau-\tau')}d\tau'  d\nu'$, where $\ast_\sigma$ denotes the twisted convolution operation.}. The cascade of twisted convolution operations in the transceiver chain gives rise to multiple integrals in the end-to-end DD domain I/O relation. The I/O relation plays a crucial role in performance evaluation and algorithm development. Closed-form expressions for the I/O relation in a compact form can aid Zak-OTFS performance evaluation and algorithm development for transceiver implementation. Motivated by this observation, in this paper, we derive closed-form expressions for the I/O relation in Zak-OTFS for sinc and Gaussian pulse shaping DD filters at the transmitter (Tx). On the receiver (Rx) side, we consider three types of DD filters, viz., $(i)$ Rx filter identical to Tx filter (referred to as ``identical filtering''), $(ii)$ Rx filter matched to the Tx filter (referred to as ``matched filtering''), and $(iii)$ Rx filter matched to both Tx filter and DD channel response (referred to as ``channel matched filtering''). 

\vspace{-1mm}
{\it Literature Survey on Zak-OTFS:}
A comprehensive overview of OTFS modulation theory and applications is provided in the recent book \cite{zak_otfs_book}. In contrast to a broader treatment in the book, this paper delves into specific aspects of Zak-OTFS, like deriving closed-form expressions for the I/O relation and noise covariance for various filtering scenarios, enabling efficient performance evaluation and comparison of different Zak-OTFS receiver structures. Apart from the two early papers on Zak-OTFS \cite{ref3},\cite{ref4}, a few other papers on the topic have appeared in the literature \cite{ref5}-\cite{zak_r1}. Identical filtering using sinc and root raised cosine (RRC) DD Tx/Rx filters in Zak-OTFS is considered in \cite{ref4} (see Eqs. (24), (25) in \cite{ref4}). But no explicit closed-form expressions for the I/O relation is given. In the absence of closed-form expressions, the integrals in the I/O relation need to be computed numerically, which is tedious and time-consuming. Matched filtering using sinc and RRC filters is considered in \cite{spread_pilot} (see Eq. (18) in \cite{spread_pilot}). Here again, no explicit closed-form I/O relation expression is given. Matched filtering is also considered in \cite{pulse_shaping} with sinc, RRC, and Gaussian Tx/Rx filters. Closed-form I/O relation expression for Gaussian filter is presented (see Theorem 1 and Eqs. (34), (35) in \cite{pulse_shaping}). Yet, closed-form expressions for other filters are not given. Channel matched filtering is considered in \cite{ref6}. For a given Tx filter, channel matched filtering at the Rx has been shown to maximize the signal-to-noise ratio (SNR) \cite{ref6}. But \cite{ref6} does not give closed-form I/O relation expressions. Other signal processing aspects of Zak-OTFS including signal detection \cite{fathima1}, LDPC-coded Zak-OTFS \cite{zak_ldpc}, iterative turbo processing based detection/channel estimation \cite{turbo_isac}, and interleaved pilot schemes \cite{int_pilot} considering sinc and RRC filters have been reported. But no closed-form I/O relation expressions for the considered systems are given.
The above works consider the Zak-OTFS framework introduced in \cite{ref3},\cite{ref4}, where quasi-periodicity is preserved using twisted convolution operation between a quasi-periodized DD signal and a DD filter which is not quasi-periodic, and these studies gave good theoretical benchmarks and insights. The works in \cite{zak_r2},\cite{zak_r1} address the issue of practical implementation of Zak-OTFS by constructing DD domain basis functions which are globally quasi-periodic and locally twisted-shifted, and using practically realizable pulse shaping filters. In our current work, we derive closed-form I/O relation expressions for the Zak-OTFS framework in \cite{ref3},\cite{ref4} with different DD filters and Tx-Rx filtering schemes, which provide good theoretical models, insights, and benchmarks.

\vspace{-1mm}
{\it New contributions:}
Our new contribution in this paper fills gaps in the previous works. We have considered all the three filtering schemes in the Zak-OTFS literature and derived closed-form expressions for all of them, providing an efficient framework for performance evaluation and comparison. Specifically, we derive discrete DD domain closed-form expressions for the I/O relation and noise covariance for identical, matched, and channel matched filtering for sinc and Gaussian filters, which have not been reported in the literature. A summary of the new contributions is highlighted in Table \ref{tab1}. For all the considered cases except for the case of sinc identical filtering, we derive exact expressions for I/O relation and noise variance in closed-form. For the sinc identical filtering case, we derive approximate closed-form expressions. Numerical results show that the approximation is accurate. Using the derived closed-form expressions, we evaluate the bit error performance of Zak-OTFS for different Tx/Rx filters and configurations. The derived closed-form expressions resulted in significant reduction in simulation run times. Our results using Vehicular-A (Veh-A) channel model with fractional DDs show that matched filtering achieves slightly better or almost same BER performance as identical filtering. Our results also show that channel matched filtering performs best in terms of BER, which is in line with the result in \cite{ref6} which shows that channel matched filtering performs best in terms of SNR.
\begin{table}[t]
\centering
\vspace{2mm}
\begin{tabular}{|l||c|c||c|c|}
\hline
&  \multicolumn{4}{|c|}{Closed-form I/O relation and noise covariance} \\ 
&  \multicolumn{4}{|c|}{expressions} \\ \cline{2-5} 
Tx/Rx filtering & \multicolumn{2}{c||}{sinc filter} & \multicolumn{2}{c|}{Gaussian filter} \\ \cline{2-5}
& \multicolumn{1}{c|}{Literature} & \multicolumn{1}{c||}{This paper} & \multicolumn{1}{c|}{Literature} & \multicolumn{1}{c|}{This paper} \\ \hline \hline
Identical            & {\color{red} $\xmark$} & {\color{blue} $\cmark$} & {\color{red} $\xmark$} & {\color{blue} $\cmark$}  \\ \hline
Matched  & {\color{red} $\xmark$}     & {\color{blue} $\cmark$} & {\color{blue} $\cmark$} \cite{pulse_shaping} & {\color{blue} $\cmark$} \\ \hline
Channel matched  & {\color{red} $\xmark$} & {\color{blue} $\cmark$} & {\color{red} $\xmark$} & {\color{blue} $\cmark$}      \\ \hline
\end{tabular}
\caption{Availability of closed-form expressions for I/O relation and noise covariance in Zak-OTFS.} 
\vspace{-5.5mm}
\label{tab1}
\end{table}

The rest of the paper is structured as follows. The Zak-OTFS system model is introduced in Sec. \ref{sec2}. Closed-form expressions for I/O relation and noise covariance for identical filtering, matched filtering, and channel matched filtering are derived in Secs. \ref{sec_idf}, \ref{sec_mf}, and \ref{sec_cmf}, respectively. Performance results obtained using the closed-form expressions are presented in Sec. \ref{sec4}. Conclusions and future work are presented in Sec. \ref{sec5}.  

\section{Zak-OTFS System Model}
\label{sec2}
\begin{figure*}
\centering    
\includegraphics[width=0.95\linewidth]{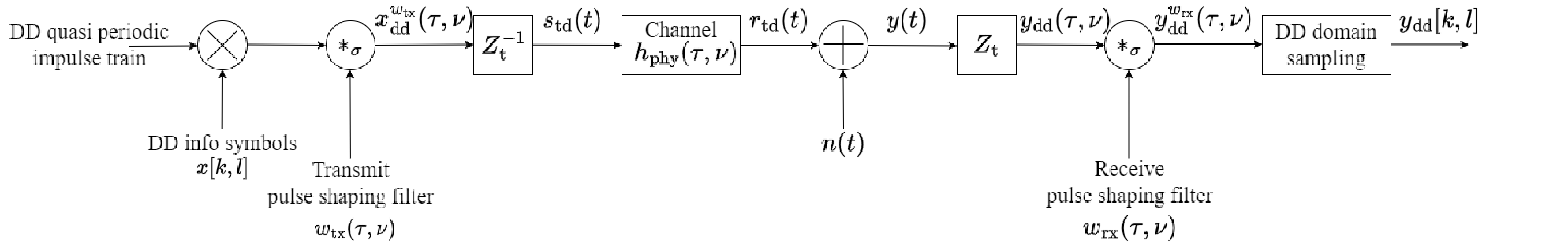}
\vspace{-0mm}
\caption{Transceiver signal processing in Zak-OTFS.}
\label{fig1} 
\vspace{-4mm}
\end{figure*}

The block diagram of the Zak-OTFS transceiver is shown in Fig. \ref{fig1}. At the transmitter, the information-bearing continuous DD domain signal is passed through the DD domain transmit filter, which makes the signal to be time- and bandwidth-limited, which is then converted into a time domain (TD) signal for transmission via inverse time-Zak transform. At the receiver, the TD signal is converted back into DD domain signal via Zak transform, followed by receive filter  and sampling operations in the DD domain for detection.

In Zak-OTFS, a pulse in DD domain, which is a quasi-periodic localized function defined by a delay period $\tau_{\mathrm{p}}$ and a Doppler period $\nu_{\mathrm{p}}=\frac{1}{\tau_{\mathrm{p}}}$, is the basic information carrier. The fundamental period in the DD domain is defined as 
\begin{equation}
\mathcal{D}_{0}= \{(\tau,\nu): 0\leq\tau<\tau_{\mathrm p}, 0\leq\nu<\nu_{\mathrm p}\},
\end{equation}
where $\tau$ and $\nu$ represent the delay and Doppler variables, respectively. The fundamental period is discretized into $M$ bins on the delay axis and $N$ bins on the Doppler axis, as 
$\{(k\frac{\tau_{{\mathrm p}}}{M},l\frac{\nu_{{\mathrm p}}}{N}) | k=0,\ldots,M-1,l=0,\ldots,N-1\}$. 
The TD Zak-OTFS frame is limited to a time duration $T=N\tau_{\mathrm p}$ and bandwidth $B=M\nu_{\mathrm p}$. In each frame, $MN$ information symbols drawn from a modulation alphabet ${\mathbb A}$, $x[k,l]\in {\mathbb A}$, $k=0,\ldots,M-1$, $l=0,\ldots,N-1$, are multiplexed on the DD domain and transmitted. The information symbol $x[k,l]$ is carried by DD domain pulse $x_{\mathrm{dd}}[k,l]$, which is a quasi-periodic function with period $M$ along the delay axis and period $N$ along the Doppler axis, i.e., for any $n,m\in\mathbb{Z}$, 
\begin{equation}
x_{\mathrm{dd}}[k+nM,l+mN]=x[k,l]e^{j2\pi n\frac{l}{N}}.
\end{equation}
These discrete DD domain signals $x_{\mathrm{dd}}[k,l]$s are supported on the information lattice 
\begin{equation}
\Lambda_{\mathrm{dd}}=
\left\{\left(k\frac{\tau_{\mathrm p}}{M},l\frac{\nu_{\mathrm p}}{N}\right) \big| k,l\in \mathbb{Z}\right\}.
\end{equation}
The continuous DD domain information symbol is given by 
\begin{equation}
x_{\mathrm{dd}}(\tau,\nu)=\sum_{k,l\in \mathbb{Z}} x_{\mathrm{dd}}[k,l] \delta\left(\tau-\frac{k\tau_{\mathrm p}}{M}\right)\delta\left(\nu-\frac{l\nu_{\mathrm p}}{N}\right),
\end{equation}
where $\delta(.)$ denotes the Dirac-delta impulse function. For any $n,m\in \mathbb{Z}$, we have
\begin{equation}
x_{\mathrm{dd}}(\tau+n\tau_{\mathrm{p}},\nu+m\nu_{\mathrm{p}})=e^{j2\pi n\nu \tau_{\mathrm{p}}}x_{\mathrm{dd}}(\tau,\nu),
\end{equation}
so that $x_{\mathrm{dd}}(\tau,\nu)$ is periodic with period $\nu_{\mathrm p}$ along the Doppler axis and quasi-periodic with period $\tau_{\mathrm p}$ along the delay axis.
The DD domain transmit signal $x_{\mathrm{dd}}^{w_{\mathrm{tx}}}(\tau,\nu)$ is given by the twisted convolution of the transmit pulse shaping filter $w_{\mathrm{tx}}(\tau,\nu)$ with $x_{\mathrm{dd}}(\tau,\nu)$ as
\begin{eqnarray}
\hspace{-6mm}
x_{\mathrm{dd}}^{w_{\mathrm{tx}}}(\tau,\nu) & \hspace{-2mm} = & \hspace{-2mm} w_{\mathrm{tx}}(\tau,\nu)*_{\sigma}x_{\mathrm{dd}}(\tau,\nu)  \nonumber \\
& \hspace{-30mm} = & \hspace{-18mm} \iint w_{\mathrm{tx}}(\tau',\nu')x_{\mathrm{dd}}(\tau-\tau',\nu-\nu') e^{j2\pi \nu'(\tau-\tau')}d\tau'd\nu',
\end{eqnarray}
where $*_{\sigma}$ denotes the twisted convolution. The transmitted TD signal $s_{\mathrm{td}}(t)$ is the TD realization of $x_{\mathrm{dd}}^{w_{\mathrm{tx}}}(\tau,\nu)$, given by
\begin{equation}
s_{\mathrm{td}}(t)=Z_{t}^{-1}\Big(x_{\mathrm{dd}}^{w_{\mathrm{tx}}}(\tau,\nu)\Big),
\end{equation}
where $Z_{t}^{-1}$ denotes the inverse time-Zak transform\footnote{Inverse time-Zak transform of a DD function $a(\tau,\nu)$ is defined as
$Z_{t}^{-1}(a(\tau,\nu)) \overset{\Delta}{=} \sqrt{\tau_{\mathrm p}} \int_0^{\nu_{\mathrm p}} a(t,\nu) d\nu$.} operation. 

Note that the pulse-shaping filter at the transmitter $w_{\mathrm{tx}}(\tau,\nu)$ is required to limit the time and bandwidth of the transmitted signal $s_{\mathrm{td}}(t)$. In the absence of the pulse-shaping filter \big(i.e., $w_{\mathrm{tx}}(\tau,\nu)=\delta(\tau)\delta(\nu)$\big), the transmitted signal has infinite duration and bandwidth. 
\textcolor{black}{The transmit signal $s_{\mathrm{td}}(t)$ passes through a doubly-selective channel to give the output signal $r_{\mathrm{td}}(t)$, where
the DD domain impulse response of the physical channel $h_{\mathrm{phy}}(\tau,\nu)$ is given by
\begin{equation}
h_{\mathrm{phy}}(\tau,\nu)=\sum_{i=1}^{P}h_{i}\delta(\tau-\tau_{i})\delta(\nu-\nu_{i}),
\label{eqn_phy}
\end{equation}
where $P$ denotes the number of DD paths, and the $i$th path has gain $h_{i}$, delay shift $\tau_{i}$, and Doppler shift $\nu_{i}$\footnote{We note that the delays ($\tau_i$s) and Dopplers ($\nu_i$s) in the channel model in (\ref{eqn_phy}) can take arbitrary values, i.e., DD values can be integer or fractional values. Also, no assumptions are made on the nature of the DD values in the I/O relation derivations in Secs. \ref{sec_idf}, \ref{sec_mf}, and \ref{sec_cmf}, and therefore these derivations are valid for the general case of fractional DDs. Consequently, the simulation results in Sec. \ref{sec4} are reported for fractional DDs.}. The received TD signal $y(t)$ at the receiver is given by
\begin{equation}
y(t)=r_{\mathrm{td}}(t)+n(t),
\end{equation}
where $n(t)$ is AWGN with variance $N_{0}$, i.e., $\mathbb{E}[n(t)n(t+t')]=N_{0}\delta(t')$. The TD signal $y(t)$ is converted to the corresponding DD domain signal $y_{\mathrm{dd}}(\tau,\nu)$ by applying Zak transform\footnote{Zak transform of a continuous TD domain signal $a(t)$ is defined as
$Z_t\left(a(t)\right) \overset{\Delta}{=} \sqrt{\tau_p} \sum_{k \in \mathbb{Z}} a(\tau + k \tau_{\mathrm p}) e^{-j2\pi\nu k\tau_{\mathrm p}}$.}, i.e.,
\begin{eqnarray}
\hspace{-6mm}
y_{\mathrm{dd}}(\tau,\nu) & \hspace{-2mm} = & \hspace{-2mm} Z_{t}(y(t))=Z_{t}(r_{\mathrm{td}}(t)+n(t)) \nonumber \\ 
& \hspace{-22mm} = & \hspace{-12mm} r_{\mathrm{dd}}(\tau,\nu)+n_{\mathrm{dd}}(\tau,\nu) \nonumber \\ 
& \hspace{-22mm} = & \hspace{-12mm} h_{\mathrm{phy}}(\tau,\nu)*_{\sigma}w_{\mathrm{tx}}(\tau,\nu)*_{\sigma}x_{\mathrm{dd}}(\tau,\nu)+n_{\mathrm{dd}}(\tau,\nu),
\end{eqnarray}
where $r_{\mathrm{dd}}(\tau,\nu)$ is the Zak transform of $r_{\mathrm{td}}(t)$, given by the twisted convolution cascade of $x_{\mathrm{dd}}(\tau,\nu)$, $w_{\mathrm{tx}}(\tau,\nu)$, and $h_{\mathrm{phy}}(\tau,\nu)$,  and $n_{\mathrm{dd}}(\tau,\nu)$ is the Zak transform of $n(t)$, i.e.,
\begin{equation}
n_{\mathrm{dd}}(\tau,\nu)=\sqrt{\tau_{\mathrm p}}\sum_{q\in \mathbb{Z}}n(\tau+q\tau_{\mathrm p})e^{-j2\pi\nu q\tau_{\mathrm p}}.
\end{equation}
The receiver filter $w_{\mathrm{rx}}(\tau,\nu)$ acts on $y_{\mathrm{dd}}(\tau,\nu)$ through twisted convolution to give the filtered output signal
\begin{eqnarray}
\hspace{-4mm} 
y_{\mathrm{dd}}^{w_{\mathrm{rx}}}(\tau,\nu) & \hspace{-2mm} = & \hspace{-2mm} w_{\mathrm{rx}}(\tau,\nu)*_{\sigma}y_{\mathrm{dd}}(\tau,\nu) \nonumber \\ 
& \hspace{-22mm} = & \hspace{-12mm} \underbrace{w_{\mathrm{rx}}(\tau,\nu)*_{\sigma}h_{\mathrm{phy}}(\tau,\nu)*_{\sigma}w_{\mathrm{tx}}(\tau,\nu)}_{\overset{\Delta}{=} \ h_{\mathrm{eff}}(\tau,\nu)}*_{\sigma}x_{\mathrm{dd}}(\tau,\nu) \nonumber \\ 
&\hspace{-22mm} & \hspace{-12mm} + \ \underbrace{w_{\mathrm{rx}}(\tau,\nu)*_{\sigma}n_{\mathrm{dd}}(\tau,\nu)}_{\overset{\Delta}{=} \ n_{\mathrm{dd}}^{w_{\mathrm{rx}}}(\tau,\nu)}, 
\label{cont1}
\end{eqnarray}
where $h_{\mathrm{eff}}(\tau,\nu)$ denotes the effective channel consisting of the twisted convolution cascade of $w_{\mathrm{tx}}(\tau,\nu),h_{\mathrm{phy}}(\tau,\nu)$, and $w_{\mathrm{rx}}(\tau,\nu)$, and $n_{\mathrm{dd}}^{w_{\mathrm{rx}}}(\tau,\nu)$ denotes the noise filtered through the Rx filter.
The DD signal $y_{\mathrm{dd}}^{w_{\mathrm{rx}}}(\tau,\nu)$ is sampled on the information lattice, resulting in the discrete quasi-periodic DD domain received signal $y_{\mathrm{dd}}[k,l]$ as
\begin{equation}
y_{\mathrm{dd}}[k,l]=y_{\mathrm{dd}}^{w_{\mathrm{rx}}}\Big(\tau=\frac{k\tau_{\mathrm p}}{M},\nu=\frac{l\nu_{\mathrm p}}{N}\Big), \ \ k,l\in\mathbb{Z},
\end{equation} 
which is given by
\begin{equation}
y_{\mathrm{dd}}[k,l]=h_{\mathrm{eff}}[k,l]*_{\sigma}x_{\mathrm{dd}}[k,l]+n_{\mathrm{dd}}[k,l],
\label{discr1}
\end{equation}
where the $*_{\sigma}$ in (\ref{discr1}) is twisted convolution in discrete DD domain, i.e., 
\vspace{-2mm}
\begin{eqnarray}
h_{\mathrm{eff}}[k,l]*_{\sigma}x_{\mathrm{dd}}[k,l]
& \hspace{-2mm} = & \hspace{-2mm} \sum_{k',l'\in\mathbb{Z}}h_{\mathrm{eff}}[k-k',l-l']x_{\mathrm{dd}}[k',l'] \nonumber \\ 
& \hspace{-2mm} & \hspace{-2mm} e^{j2\pi\frac{k'(l-l')}{MN}},
\end{eqnarray}
where the effective channel filter $h_{\mathrm{eff}}[k,l]$ and filtered noise samples $n_{\mathrm{dd}}[k,l]$ are given by
\vspace{-2mm}
\begin{align}
h_{\mathrm{eff}}[k,l]=h_{\mathrm{eff}}\Big(\tau=\frac{k\tau_{p}}{M},\nu=\frac{l\nu_{p}}{N}\Big), \label{discr2} \\ 
n_{\mathrm{dd}}[k,l]=n_{dd}^{w_{\mathrm{rx}}}\Big(\tau=\frac{k\tau_{p}}{M},\nu=\frac{l\nu_{p}}{N}\Big).
\label{discr3}
\end{align}
}

\vspace{-4mm}
Owing to the quasi-periodicity in the DD domain, it is sufficient to consider the received samples $y_{\mathrm{dd}}[k,l]$ within the fundamental period $\mathcal{D}_0$. We write the $y_{\mathrm{dd}}[k,l]$ samples as a vector and the end-to-end DD domain I/O relation in matrix-vector form as
\begin{equation}
\mathbf{y}=\mathbf{Hx}+\mathbf{n},
\label{sys_mod}
\end{equation}
where $\mathbf{x,y,n} \in\mathbb{C}^{MN\times 1}$, such that their $(kN+l+1)$th entries are given by $x_{kN+l+1}=x_{\mathrm{dd}}[k,l]$, $y_{kN+l+1}=y_{\mathrm{dd}}[k,l]$, $n_{kN+l+1}=n_{\mathrm{dd}}[k,l]$, and $\mathbf{H}\in\mathbb{C}^{MN\times MN}$ is the channel matrix such that
\vspace{-1mm}
\begin{eqnarray}
\mathbf{H}[k'N+l'+1,kN+l+1] & \hspace{-2mm} = & \hspace{-2mm} \sum_{m,n\in\mathbb{Z}}h_{\mathrm{eff}}[k'-k-nM, \nonumber \\
& \hspace{-45mm} & \hspace{-35mm} l'-l-mN]e^{j2\pi nl/N}e^{j2\pi\frac{(l'-l-mN)(k+nM)}{MN}},
\label{eqn_channel_matrix}
\vspace{-4mm}
\end{eqnarray}
where $k',k=0,\ldots,M-1$, $l',l=0,\ldots,N-1$. 

The system model in (\ref{sys_mod}) allows performance evaluation of Zak-OTFS for different choices of Tx/Rx filters and can aid Zak-OTFS transceiver algorithms development. However, the expressions for $h_{\mathrm{eff}}[k,l]$ and covariance of $n_{\mathrm{dd}}[k,l]$ involve multiple integrals due to twisted convolution operations (see (\ref{cont1})), the computation of which is tedious and consumes substantial run times. We alleviate this by deriving closed-form expressions for
$h_{\mathrm{eff}}[k,l]$ and covariances of $n_{\mathrm{dd}}[k,l]$ for different Tx/Rx filter configurations, which can  be computed in closed-form to obtain the channel matrix ${\bf H}$ and the noise vector ${\bf n}$ in (\ref{sys_mod}). This needs the derivation of the cascade of twisted convolution in (\ref{cont1}) among Tx filter $w_{\mathrm{tx}}(\tau,\nu)$, physical channel response $h_{\mathrm{phy}}(\tau,\nu)$, and Rx filter $w_{\mathrm{rx}}(\tau,\nu)$.
In Secs. \ref{sec_idf}, \ref{sec_mf}, and \ref{sec_cmf}, we carry out the above derivations for sinc and Gaussian Tx filters and the following choices of Rx filter.
\vspace{-3mm}
\subsection{Identical Filtering}
In this case, a Rx filter that is identical to the Tx filter is used, i.e., the Rx filter is chosen to be \cite{ref4}
\begin{equation}
w_{\mathrm{rx}}(\tau,\nu)=w_{\mathrm{tx}}(\tau,\nu).
\label{id_filtering}
\end{equation}
\vspace{-9mm}
\subsection{Matched Filtering}
Here, a Rx filter that is matched to the Tx filter is used, i.e., the Rx filter is chosen to be \cite{spread_pilot}, \cite{pulse_shaping} 
\begin{equation}
w_{\mathrm{rx}}(\tau,\nu)=w_{\mathrm{tx}}^{\dagger}(\tau,\nu)=w_{\mathrm{tx}}^{*}(-\tau,-\nu) e^{j2\pi \nu\tau}.
\label{mat_filtering}
\end{equation}
\vspace{-9mm}
\subsection{Channel Matched Filtering}
In this case, a Rx filter that is matched to the cascade of the Tx filter and the physical channel response is used, i.e., the Rx filter is chosen to be \cite{ref6}\footnote{We note that the channel matched filter is motivated by the work in \cite{ref6} (Theorem 5), where it has been shown that, for any arbitrary Tx filter $w_\text{tx}(\tau,\nu)$, the optimal Rx filter that maximizes the SNR is the one that is matched to the physical channel $h_\text{phy}(\tau,\nu)$ and the Tx filter $w_\text{tx}(\tau,\nu)$, which we have termed as the channel matched filtering in this paper. Therefore, deriving closed-form expressions for the system with this filter and evaluating its BER performance can serve as a performance benchmark for comparing the performance of different filtering schemes.}
\begin{equation}
w_{\mathrm{rx}}(\tau,\nu) \hspace{-0.5mm} =  \hspace{-0.5mm} (h_{\mathrm{phy}}(\tau,\nu)*_{\sigma}w_{\mathrm{tx}}(\tau,\nu))^{\dagger} \hspace{-0.5mm} = \hspace{-0.5mm} w_{\mathrm{tx}}^{\dagger}(\tau,\nu)*_{\sigma}h_{\mathrm{phy}}^{\dagger}(\tau,\nu). 
\label{cmat_filtering}
\end{equation}
The Tx filter is considered to be of the form $w_{\mathrm{tx}}(\tau,\nu)=w_1(\tau)w_2(\nu)$ \cite{ref3},\cite{ref4}, so that the sinc Tx filter is of the form
\begin{equation}
w_{\mathrm{tx}}(\tau,\nu)=\sqrt{BT} \hspace{0.5mm} \mathrm{sinc}(B\tau) \hspace{0.5mm} \mathrm{sinc}(T\nu),
\end{equation}
where $w_1(\tau)=\sqrt{B} \hspace{0.3mm} \mathrm{sinc}(B\tau)$ and $w_2(\nu)= \sqrt{T} \hspace{0.3mm} \mathrm{sinc}(T\nu)$,
and the Gaussian Tx filter is of the form
\begin{equation}
\hspace{-2mm}
w_{\mathrm{tx}}(\tau,\nu) =  
\Big(\frac{2\alpha_{\tau}B^{2}}{\pi}\Big)^{\frac{1}{4}}e^{-\alpha_{\tau}B^{2}\tau^{2}}
\Big(\frac{2\alpha_{\nu}T^{2}}{\pi}\Big)^{\frac{1}{4}}e^{-\alpha_{\nu}T^{2}\nu^{2}}
\hspace{-1mm}.
\end{equation}
Since Gaussian pulse has infinite support, we work with time interval $T'$ where 99$\%$ of the Zak-OTFS frame energy is localized in TD, and with bandwidth $B'$ where 99$\%$ of the frame energy is localized in FD. We can configure Gaussian pulse by adjusting the parameters $\alpha_{\tau}$ and $\alpha_{\nu}$. 
No time/bandwidth expansion ($T'=T$, $B'=B$) corresponds to $\alpha_{\tau}=\alpha_{\nu}=1.584$. 

\section{Closed-form Expressions for Identical Filtering}
\label{sec_idf}
In this section, we derive closed-form expressions for the I/O relation and noise covariance for 
identical filtering with sinc and Gaussian Tx filters. The effective channel in the continuous DD domain for identical filtering, i.e., for $w_{\mathrm{rx}}=w_{\mathrm{tx}}$, can be written as

\vspace{-4mm}
{\small
\begin{eqnarray}
\hspace{-4mm}
h_{\mathrm{eff}}(\tau,\nu) & \hspace{-2mm} = & \hspace{-2mm}  w_{\mathrm{rx}}(\tau,\nu)*_{\sigma}h_{\mathrm{phy}}(\tau,\nu)*_{\sigma}w_{\mathrm{tx}}(\tau,\nu) \nonumber \\
& \hspace{-27mm} = & \hspace{-15mm} w_{\mathrm{rx}}(\tau,\nu)*_{\sigma}\left(\sum_{i=1}^{P}h_{i}\delta(\tau-\tau_{i})\delta(\nu-\nu_{i})\right)*_{\sigma}w_{1}(\tau)w_{2}(\nu) \nonumber \\ 
& \hspace{-27mm} = & \hspace{-15mm} w_{1}(\tau)w_{2}(\nu)*_{\sigma}\left(\sum_{i=1}^{P}h_{i}w_{1}(\tau-\tau_{i})w_{2}(\nu-\nu_{i})e^{j2\pi\nu_{i}(\tau-\tau_{i})}\right) \nonumber \\
& \hspace{-27mm} = & \hspace{-15mm} \sum_{i=1}^{P}h_{i}e^{j2\pi\nu_{i}(\tau-\tau_{i})}\bigg(\int w_{1}(\tau')w_{1}(\tau-\tau'-\tau_{i})e^{-j2\pi\nu_{i}\tau'} \nonumber \\ 
& \hspace{-27mm} & \hspace{-15mm} \underbrace{\left(\int w_{2}(\nu')w_{2}(\nu-\nu'-\nu_{i})e^{j2\pi\nu'(\tau-\tau')}d\nu'\right)}_{\overset{\Delta}{=}I_{i}^{(1)}(\tau,\tau',\nu)}d\tau'\bigg).
\label{eqn:identical_channel}
\end{eqnarray}
}
The continuous DD domain noise at the output is
\begin{eqnarray}
\hspace{2mm}
n_{\mathrm{dd}}^{w_{\mathrm{rx}}}(\tau,\nu) & \hspace{-66mm} = & \hspace{-66mm} w_{\mathrm{rx}}(\tau,\nu)*_{\sigma} n_{\mathrm{dd}}(\tau,\nu) \nonumber \\ 
& \hspace{-15mm} = \hspace{-0mm} w_{1}(\tau)w_{2}(\nu)*_{\sigma}\left(\sqrt{\tau_{p}}\sum_{q\in \mathbb{Z}}n(\tau+q\tau_{p})e^{-j2\pi\nu q\tau_{p}}\right) \nonumber \\ 
& \hspace{-15mm} = \hspace{-0mm} 
 \sqrt{\tau_{p}}\sum_{q=-\infty}^{\infty}e^{-j2\pi\nu q\tau_{p}}\bigg(\int
 w_{1}(\tau')n(\tau-\tau'+q\tau_{p}) \nonumber \\
& \hspace{-77mm} & \hspace{-77mm}  \underbrace{\left(\int w_{2}(\nu')e^{j2\pi\nu'(\tau-\tau'+q\tau_{p})}d\nu'\right)}_{\overset{\Delta}{=}I_{q}^{(2)}(\tau,\tau')}d\tau'\bigg). 
\label{eqn:noise_identical}
\end{eqnarray}

\subsection{$h_{\mathrm{eff}}[k,l]$ for sinc filter} 
Integration of the inner integral defined as $I_{i}^{(1)}(\tau,\tau',\nu)$, $ 1\leq i\leq P$, in (\ref{eqn:identical_channel}) gives
\begin{eqnarray}
\hspace{-8mm}
I_{i}^{(1)}(\tau,\tau',\nu) & \hspace{-2mm} = & \hspace{-2mm} e^{j\pi(\tau-\tau')(\nu-\nu_{i})}\left(\frac{T-|\tau-\tau'|}{T}\right) \nonumber    \\ 
& \hspace{-15mm}  & \hspace{-15mm} \mathrm{sinc}((T-|\tau-\tau'|)(\nu-\nu_{i}))\mathbbm{1}_{\{-T<\tau-\tau'<T\}},
\label{eqn:Ii_identical}
\vspace{-2mm}
\end{eqnarray}
where $\mathbbm{1}_{\{.\}}$ denotes the indicator function. Substituting (\ref{eqn:Ii_identical}) in (\ref{eqn:identical_channel}) gives
\vspace{-4mm}
\begin{eqnarray}
\hspace{-3mm}
h_{\mathrm{eff}}(\tau,\nu) & \hspace{-2mm} =  & \hspace{-2mm} \left(\frac{B}{T}\right)\sum_{i=1}^{P}h_{i}e^{-j2\pi\tau_{i}\nu_{i}}\bigg(\int e^{j\pi(\tau-\tau')(\nu+\nu_{i})} \nonumber \\
&  \hspace{-8mm} &  \hspace{-8mm} \mathrm{sinc}(B\tau')\mathrm{sinc}(B(\tau-\tau'-\tau_{i}))(T-|\tau-\tau'|) \nonumber \\ 
& \hspace{-8mm} & \hspace{-8mm} \mathrm{sinc}((T-|\tau-\tau'|)(\nu-\nu_{i}))\mathbbm{1}_{\{-T< \tau-\tau'< T\}} d \tau'\bigg)\nonumber \\ 
& \hspace{-30mm} = & \hspace{-16mm} \left(\frac{B}{T}\right)\sum_{i=1}^{P}h_{i}e^{-j2\pi\tau_{i}\nu_{i}}\bigg(\int_{-T}^{T}e^{-j\pi x(\nu+\nu_{i})} \mathrm{sinc}(B(x+\tau)) \nonumber \\ 
& \hspace{-18mm} & \hspace{-18mm} \mathrm{sinc}(B(x+\tau_{i}))(T-|x|)\mathrm{sinc}((T-|x|)(\nu-\nu_{i}))dx\bigg),
\label{eqn:Exact_identical_channel}
\vspace{-4mm}
\end{eqnarray}
where the last step is by the substitution $x=\tau'-\tau$, and $h_{\mathrm{eff}}[k,l]$ is obtained by sampling $h_{\mathrm{eff}}(\tau,\nu)$ on the information lattice (see (\ref{discr2})). Analytical simplification of the integral in (\ref{eqn:Exact_identical_channel}) into an exact closed-form is difficult, and hence we obtain an approximate closed-form expression for $h_{\mathrm{eff}}[k,l]$ as follows.
\begin{thm}
For identical filtering with sinc filter, 
the effective channel $h_{\mathrm{eff}}[k,l]$ in approximate closed-from is given by
\begin{eqnarray}
h_{\mathrm{eff}}[k,l] & \hspace{-2mm} \approx & \hspace{-2mm} \left(\frac{B}{2}\right)\sum_{i=1}^{P}h_{i}e^{-j2\pi\tau_i\nu_i}\mathrm{sinc}\left(T\left(\frac{l\nu_{p}}{N}-\nu_{i}\right)\right) \nonumber \\
& \hspace{-2mm} & \hspace{-2mm} \left(P_{i,k}\left(\frac{l\nu_{p}}{N}\right)+P_{i,k}(\nu_{i})\right),
\label{eqn:sinc_identical_closed}
\end{eqnarray}
where the function $P_{i,k}(f)$ is given by
\begin{align}
P_{i,k}(f)=&e^{j\pi f(\frac{k\tau_p}{M}+\tau_{i})}\left(\frac{B-|f|}{B^{2}}\right) \nonumber \\
&\mathrm{sinc}\left((B-|f|)\left(\frac{k\tau_p}{M}-\tau_{i}\right)\right)\mathbbm{1}_{\{-B<f<B\}}.
\label{eqn:P_ki_discrete}
\end{align}
\end{thm}
\vspace{-1mm}
\begin{IEEEproof}
See Appendix \ref{appendix_a1}.
\end{IEEEproof}

\vspace{-1mm} 
\subsection{Noise covariance for sinc filter}
For sinc filter at the transmitter, the integral $I_{q}^{(2)}(\tau,\tau')$ in (\ref{eqn:noise_identical}) is simplified as
\vspace{-2mm}
\begin{eqnarray}
I_{q}^{(2)}(\tau,\tau') & \hspace{-2mm} = & \hspace{-2mm} \sqrt{T} \int \mathrm{sinc}(T\nu')e^{j2\pi\nu'(\tau-\tau'+q\tau_{p})}d\nu' \nonumber \\
& \hspace{-2mm} = & \hspace{-2mm} \frac{1}{\sqrt{T}}\mathrm{rect}\left(\frac{\tau-\tau'+q\tau_{p}}{T}\right),
\label{eqn:noise_integral}
\end{eqnarray}
where the $\mathrm{rect}(.)$ function is defined as
\cite{rect_ref_book1}
\begin{equation}
\mathrm{rect}\left(\frac{t}{T}\right)=\begin{cases}
    0, \quad  |t|> \frac{T}{2} \\
    \frac{1}{2}, \quad  |t|= \frac{T}{2} \\
    1, \quad  |t|< \frac{T}{2}.    
    \end{cases}
\end{equation}
Substituting (\ref{eqn:noise_integral}) in (\ref{eqn:noise_identical}) gives
\begin{eqnarray}
\hspace{-6mm}
n_{\mathrm{dd}}^{w_{\mathrm{rx}}}(\tau,\nu) & \hspace{-2mm} = & \hspace{-2mm} \sqrt{\frac{B\tau_{p}}{T}}\sum_{q=-\infty}^{\infty}e^{-j2\pi\nu q\tau_{p}}\Bigg(\int \mathrm{sinc}(B\tau') \nonumber
\\ 
& \hspace{-2mm}  & \hspace{-2mm} n(\tau-\tau'+q\tau_{p}) \mathrm{rect}\left(\frac{\tau-\tau'+q\tau_{p}}{T}\right)d\tau'\Bigg) \nonumber \\ 
& \hspace{-10mm}  = & \hspace{-5mm}  \sqrt{\frac{B\tau_{p}}{T}}\sum_{q=-\infty}^{\infty}e^{-j2\pi\nu q\tau_{p}} \nonumber \\ 
& \hspace{-10mm}  & \hspace{-5mm} \underbrace{\left(\int_{-\frac{T}{2}}^{\frac{T}{2}}\mathrm{sinc}(B(\tau+q\tau_{p}+x))n(-x)dx\right)}_{\overset{\Delta}{=}f(\tau+q\tau_{p})},
\label{eqn:filtered_noise_identical}
\end{eqnarray}
where the last step comes via the substitution $x=\tau'-\tau-q\tau_{p}$. 

Sampling (\ref{eqn:filtered_noise_identical}) on $\Lambda_{\mathrm{dd}}=\{(k\frac{\tau_{p}}{M},l\frac{\nu_{p}}{N}) | k,l\in \mathbb{Z}\}$ gives 
\begin{equation}
n_{\mathrm{dd}}[k,l]=\sqrt{\frac{B\tau_{p}}{T}}\sum_{q=-\infty}^{\infty}e^{-j2\pi ql/N}f\left(\frac{k\tau_{p}}{M}+q\tau_{p}\right).
\label{eqn:sampled_noise_identical}
\end{equation}
From (\ref{eqn:sampled_noise_identical}), we can write the noise covariance as
\begin{eqnarray}
\hspace{-7mm}
\mathbb{E}[n_{\mathrm{dd}}[k_{1},l_{1}]n_{\mathrm{dd}}^{*}[k_{2},l_{2}]] & \hspace{-2mm} = & \hspace{-2mm} \left(\frac{B\tau_{p}}{T}\right)\sum_{q_{1}=-\infty}^{\infty}\sum_{q_{2}=-\infty}^{\infty} \nonumber \\ 
& \hspace{-41mm} & \hspace{-41mm} e^{j2\pi\frac{q_{2}l_{2}-q_{1}l_{1}}{N}}\underbrace{\mathbb{E}\bigg[f\left(\frac{k_{1}\tau_{p}}{M}+q_{1}\tau_{p}\right)f^{*}\left(\frac{k_{2}\tau_{p}}{M}+q_{2}\tau_{p}\right)\bigg]}_{\overset{\Delta}{=}\mathbb{E}_{\{k_{1},k_{2},q_{1},q_{2}\}}},
\label{eqn:noise_covariance_1}
\end{eqnarray}
where the term $\mathbb{E}_{\{k_{1},k_{2},q_{1},q_{2}\}}$ is given by
\begin{eqnarray}
\mathbb{E}_{\{k_{1},k_{2},q_{1},q_{2}\}} & \hspace{-3mm} = & \hspace{-3mm} 
\int_{-\frac{T}{2}}^{\frac{T}{2}}\int_{-\frac{T}{2}}^{\frac{T}{2}}\mathrm{sinc}\left(B\left(\frac{k_{1}\tau_{p}}{M}+q_{1}\tau_{p}+x_{1}\right) \right) \nonumber \\ 
& \hspace{-25mm} & \hspace{-25mm} \mathrm{sinc}\left(B\left(\frac{k_{2}\tau_{p}}{M}+q_{2}\tau_{p}+x_{2}\right)\right)\underbrace{\mathbb{E}[n(-x_{1})n^{*}(-x_{2})]}_{=N_{0}\delta(x_{2}-x_{1})}dx_{1}dx_{2} 
\nonumber \\ 
& \hspace{-37mm} = & \hspace{-20mm} N_{0}\bigg[\int_{-\frac{T}{2}}^{\frac{T}{2}}\hspace{-2mm} \mathrm{sinc}\left(B\left(\frac{k_{1}\tau_{p}}{M}+q_{1}\tau_{p}+x\right)\right)\mathrm{sinc}\bigg(B\bigg(\frac{k_{2}\tau_{p}}{M} \nonumber \\ 
& \hspace{-37mm} & \hspace{-20mm} +q_{2}\tau_{p}+x\bigg)\bigg)dx\bigg].
\label{eqn:noise_covariance_2}
\end{eqnarray}
Substituting (\ref{eqn:noise_covariance_2}) in (\ref{eqn:noise_covariance_1}), we get 
\begin{eqnarray}
\mathbb{E}[n_{\mathrm{dd}}[k_{1},l_{1}]n_{\mathrm{dd}}^{*}[k_{2},l_{2}]] & \hspace{-2mm} = & \hspace{-2mm} N_{0}\left(\frac{B\tau_{p}}{T}\right)\sum_{q_{1}=-\infty}^{\infty}\sum_{q_{2}=-\infty}^{\infty} \nonumber    \\ 
& \hspace{-37mm} & \hspace{-37mm} e^{j2\pi\frac{q_{2}l_{2}-q_{1}l_{1}}{N}}\bigg(\int_{-\frac{T}{2}}^{\frac{T}{2}}\hspace{-2mm} \mathrm{sinc}\left(B\left(\frac{k_{1}\tau_{p}}{M}+q_{1}\tau_{p}+x\right)\right) \nonumber \\ 
& \hspace{-37mm} & \hspace{-37mm}  \mathrm{sinc}\left(B\left(\frac{k_{2}\tau_{p}}{M}+q_{2}\tau_{p}+x\right)\right)dx\bigg).
\label{eqn:noise_covariance_identical}
\end{eqnarray}
For large $M$ and $N$ (consequently, large $B$ and $T$), the integral in (\ref{eqn:noise_covariance_identical}) can be approximated as 
\begin{eqnarray}
\int_{-\frac{T}{2}}^{\frac{T}{2}}\mathrm{sinc}\bigg(B\bigg(\frac{k_{1}\tau_{p}}{M}+q_{1}\tau_{p}+x\bigg)\bigg) & & \nonumber \\
\mathrm{sinc}\bigg(B\bigg(\frac{k_{2}\tau_{p}}{M}+q_{2}\tau_{p}+x\bigg)\bigg)dx &  & \nonumber \\ 
& \hspace{-109mm} \approx & \hspace{-56.5mm} \begin{cases}
    0, \forall  k_{1}\neq k_{2};\ 0\leq k_{1},k_{2}\leq M-1, \\
    0, \forall  q_{1}\neq q_{2};\ q_{1},q_{2}\in \mathbb{Z} \\
    \left(\frac{1}{B}\right), \forall k_{1}=k_{2},\ q_{1}=q_{2},\ q_{1}\tau_{p}+\frac{k_{1}\tau_{p}}{M}\in\left(-\frac{T}{2},\frac{T}{2}\right) \\
    \left(\frac{1}{2B}\right), \forall  k_{1}=k_{2},\ q_{1}=q_{2},\ q_{1}\tau_{p}+\frac{k_{1}\tau_{p}}{M}\in\big\{-\frac{T}{2},\frac{T}{2}\big\} \\
    0,\mathrm{otherwise}.
    \end{cases}
\label{eqn:noise_integral_closed}
\end{eqnarray}
Substituting (\ref{eqn:noise_integral_closed}) in (\ref{eqn:noise_covariance_identical}) and subsequently solving the summation (which is finite sum due to the limit in the range of $q_1$ and $q_2$) 
will give the noise covariance expression, i.e., 
for all $k_{1},k_{2}=0,1,\ldots,M-1$, $l_{1},l_{2}=0,1,\ldots,N-1$, the $(k_{1}N+l_{1}+1,k_{2}N+l_{2}+1)$th element of the noise covariance matrix 
is given by
\begin{equation}
\mathbb{E}[n_{\mathrm{dd}}[k_{1},l_{1}]n_{\mathrm{dd}}^{*}[k_{2},l_{2}]]\approx\begin{cases}
    N_{0}, \forall k_{1}=k_{2},l_{1}=l_{2} \\
    0, \mathrm{otherwise}.    
    \end{cases}
\label{eqn:noise_approx_closed}
\end{equation}

\subsection{$h_{\mathrm{eff}}[k,l]$ for Gaussian filter} 
Integration of $I_{i}^{\textcolor{black}{(1)}}(\tau,\tau',\nu)$ in (\ref{eqn:identical_channel}) for Gaussian filter gives
\begin{equation}
I_{i}^{(1)}(\tau,\tau',\nu) \hspace{-0mm} =  \hspace{-0mm} e^{-\frac{\alpha_{\nu}T^{2}}{2}\left((\nu-\nu_{i})^{2}-j2\pi\frac{(\nu-\nu_{i})(\tau-\tau')}{\alpha_{\nu}T^{2}}+\left(\frac{\pi (\tau-\tau')}{\alpha_{\nu}T^{2}}\right)^{2}\right)}.
\label{eqn:Ii_identical_gaussian}
\end{equation}
Substituting (\ref{eqn:Ii_identical_gaussian}) in (\ref{eqn:identical_channel}) gives
\begin{eqnarray}
\hspace{-8mm} 
h_{\mathrm{eff}}(\tau,\nu) & \hspace{-2mm} = & \hspace{-2mm} \left(\frac{2\alpha_{\tau}B^{2}}{\pi}\right)^{\frac{1}{2}}\sum_{i=1}^{P}h_{i}e^{j2\pi\nu_{i}(\tau-\tau_{i})} \nonumber \\ 
& \hspace{-16mm} & \hspace{-16mm} \bigg(\int e^{-\alpha_{\tau}B^{2}\tau_{1}^{2}}e^{-\alpha_{\tau}B^{2}(\tau-\tau_{1}-\tau_{i})^{2}}e^{-j2\pi\nu_{i}\tau_{1}} \nonumber \\ 
& \hspace{-16mm} & \hspace{-16mm}  e^{-\frac{\alpha_{\nu}T^{2}}{2}\left((\nu-\nu_{i})^{2}-j2\pi\frac{(\nu-\nu_{i})(\tau-\tau_{1})}{\alpha_{\nu}T^{2}}+\left(\frac{\pi (\tau-\tau_{1})}{\alpha_{\nu}T^{2}}\right)^{2}\right)}d\tau_{1}\bigg).
\label{eqn:gauss_identical_channel}
\end{eqnarray}
The integral in (\ref{eqn:gauss_identical_channel}) can be solved analytically to obtain a 
closed-form expression for $h_{\mathrm{eff}}(\tau,\nu)$, which can be sampled on the information lattice to obtain $h_{\mathrm{eff}}[k,l]$ in exact closed-form. 
The following theorem states the result.
\begin{thm}
For identical filtering with Gaussian filter, the effective channel $h_{\mathrm{eff}}[k,l]$ in exact closed-form is given by
\begin{equation}
h_{\mathrm{eff}}[k,l] = \left(\frac{2\alpha_{\tau}B^{2}}{2\alpha_{\tau}B^{2}+\frac{\pi^{2}}{2\alpha_{\nu}T^{2}}}\right)^{\frac{1}{2}}\sum_{i=1}^{P}h_{i}
    e^{-g_{i}[k,l]},
\label{eqn:gauss_identical_discrete}
\end{equation}
where the function $g_{i}[k,l]$, $\forall \ 1\leq i\leq P$ is given by 
\begin{eqnarray}
\hspace{-4mm}
g_{i}[k,l] & \hspace{-2mm} = & \hspace{-2mm}  \alpha_{\tau}B^{2}\left(\frac{k^{2}\tau_{p}^{2}}{M^{2}}+\tau_{i}^{2}\right)+j2\pi\nu_{i}\tau_{i}+\frac{\alpha_{\nu}T^{2}}{2}\bigg(\frac{l\nu_{p}}{N}
\nonumber    \\ 
& \hspace{-12mm} & \hspace{-12mm} -\nu_{i}\bigg)^{2} -\frac{\left(2\alpha_{\tau}B^{2}\left(\frac{k\tau_{p}}{M}+\tau_{i}\right)+j\pi\left(\frac{l\nu_{p}}{N}+\nu_{i}\right)\right)^{2}}{4\left(2\alpha_{\tau}B^{2}+\frac{\pi^{2}}{2\alpha_{\nu}T^{2}}\right)}.
\label{eqn:gi_discrete}
\end{eqnarray}
\end{thm}
\begin{IEEEproof}
See Appendix \ref{appendix_b1}
\end{IEEEproof}
 
\vspace{-4mm}
\subsection{Noise covariance for Gaussian filter}
Integration of the inner integral 
$I_{q}^{(2)}(\tau,\tau')$ in (\ref{eqn:noise_identical}) for Gaussian filter gives
\begin{equation}
I_{q}^{(2)}(\tau,\tau')=\left(\frac{2\pi}{\alpha_{\nu}T^{2}}\right)^{\frac{1}{4}}e^{-\frac{\pi^{2}(\tau-\tau'+q\tau_{p})^{2}}{\alpha_{\nu}T^{2}}}.
\label{eqn:gaussian_noise_Iq}
\end{equation}
Substituting (\ref{eqn:gaussian_noise_Iq}) in (\ref{eqn:noise_identical}), we obtain
\begin{eqnarray}
n_{\mathrm{dd}}^{w_{\mathrm{rx}}}(\tau,\nu) & \hspace{-2mm} = & \hspace{-2mm} \sqrt{\tau_{p}}\left(\frac{2\pi}{\alpha_{\nu}T^{2}}\right)^{\frac{1}{4}}\sum_{q=-\infty}^{\infty}e^{-j2\pi\nu q\tau_{p}} \nonumber \\
& \hspace{-8mm} & \hspace{-8mm} \bigg(\int w_{1}(\tau')n(\tau-\tau'+q\tau_{p})e^{-\frac{\pi^{2}(\tau-\tau'+q\tau_{p})^{2}}{\alpha_{\nu}T^{2}}}d\tau'\bigg) \nonumber \\ 
& \hspace{-2mm} = & \hspace{-2mm} \sqrt{\frac{2B\tau_{p}}{T}}\left(\frac{\alpha_{\tau}}{\alpha_{\nu}}\right)^{\frac{1}{4}}\sum_{q=-\infty}^{\infty}e^{-j2\pi\nu q\tau_{p}} \nonumber \\ 
& \hspace{-20mm} & \hspace{-20mm} \underbrace{\left(\int e^{-\alpha_{\tau}B^{2}\tau_{1}^{2}}e^{-\frac{\pi^{2}(\tau-\tau_{1}+q\tau_{p})^{2}}{\alpha_{\nu}T^{2}}}n(\tau-\tau_{1}+q\tau_{p})d\tau_{1}\right)}_{\overset{\Delta}{=}f(\tau+q\tau_{p})}.
\label{eqn:gaussian_noise_output}
\end{eqnarray}
Sampling (\ref{eqn:gaussian_noise_output}) on 
$\Lambda_{\mathrm{dd}}=\{(k\frac{\tau_{p}}{M},l\frac{\nu_{p}}{N}) | k,l\in \mathbb{Z}\}$ gives
\begin{equation}
n_{\mathrm{dd}}[k,l]=\sqrt{\frac{2B\tau_{p}}{T}}\left(\frac{\alpha_{\tau}}{\alpha_{\nu}}\right)^{\frac{1}{4}}\hspace{-2mm}\sum_{q=-\infty}^{\infty} \hspace{-2mm} e^{-j2\pi ql/N}f\left(\frac{k\tau_{p}}{M}+q\tau_{p}\right).
\label{eqn:gaussian_noise_covariance1}
\end{equation}
From (\ref{eqn:gaussian_noise_covariance1}), the noise covariance is given by
\begin{eqnarray}
\hspace{-3mm}
\mathbb{E}[n_{\mathrm{dd}}[k_{1},l_{1}]n_{\mathrm{dd}}^{*}[k_{2},l_{2}]] & \hspace{-2mm} = & \hspace{-2mm}  \left(\frac{2B\tau_{p}}{T}\right)\sqrt{\frac{\alpha_{\tau}}{\alpha_{\nu}}}\sum_{q_{1}=-\infty}^{\infty}\sum_{q_{2}=-\infty}^{\infty} \nonumber \\ 
& \hspace{-40mm} & \hspace{-40mm} e^{j2\pi\frac{q_{2}l_{2}-q_{1}l_{1}}{N}}\underbrace{\mathbb{E}\bigg[f\hspace{-1mm}\left(\frac{k_{1}\tau_{p}}{M}+q_{1}\tau_{p}\right)\hspace{-0mm} f^{*} \hspace{-1mm}\left(\frac{k_{2}\tau_{p}}{M}+q_{2}\tau_{p}\right)\bigg]}_{\overset{\Delta}{=}\mathbb{E}_{\{k_{1},k_{2},q_{1},q_{2}\}}}\hspace{-0.5mm}.
\label{eqn:gaussian_noise_sampled}
\end{eqnarray}
We can analytically solve the term $\mathbb{E}_{\{k_{1},k_{2},q_{1},q_{2}\}}$ in (\ref{eqn:gaussian_noise_sampled}) to obtain the following result in exact closed-form.
\begin{thm}
For identical filtering with Gaussian filter,
for all $k_{1},k_{2}=0,1,\ldots,M-1$, $l_{1},l_{2}=0,1,\ldots,N-1,$ the $(k_{1}N+l_{1}+1,k_{2}N+l_{2}+1)$th element of the noise covariance matrix  
is given by
\begin{eqnarray}
\mathbb{E}[n_{\mathrm{dd}}[k_{1},l_{1}]n_{\mathrm{dd}}^{*}[k_{2},l_{2}]] & \hspace{-3mm} = & \hspace{-3mm} N_{0}\hspace{-0.5mm}\left(\frac{2B\tau_{p}}{T}\right)\hspace{-1mm}\left(\frac{\pi\alpha_{\tau}}{2\alpha_{\tau}\alpha_{\nu}B^{2}+\frac{2\pi^{2}}{T^{2}}}\right)^{\hspace{0mm}\frac{1}{2}} \nonumber \\
& \hspace{-25mm} & \hspace{-25mm} \sum_{q_{1}=-\infty}^{\infty}\sum_{q_{2}=-\infty}^{\infty}e^{j2\pi\frac{q_{2}l_{2}-q_{1}l_{1}}{N}}\hspace{-2mm} e^{-\frac{g[k_{1},k_{2},q_{1},q_{2}]}{\left(2\alpha_{\tau}B^{2}+\frac{2\pi^{2}}{\alpha_{\nu}T^{2}}\right)}},
\label{eqn:gaussian_noise_covariance}
\end{eqnarray}
where the function $g[k_{1},k_{2},q_{1},q_{2}]$, $\forall \ 0\leq k_{1},k_{2}\leq M-1$ and $q_{1},q_{2}\in\mathbb{Z}$, is given by
\begin{eqnarray}           
\hspace{-4mm}
g[k_{1},k_{2},q_{1},q_{2}] & \hspace{-2mm} = & \hspace{-2mm} \left(\alpha_{\tau}B^{2}\right)^{2}\left(\frac{k_{2}-k_{1}}{M}+(q_{2}-q_{1})\right)^{2}\tau_{p}^{2} \nonumber \\ 
& \hspace{-22mm} & \hspace{-22mm} + \ 2\pi^{2}\frac{\alpha_{\tau}B^{2}}{\alpha_{\nu}T^{2}}\left(\left(\frac{k_{1}}{M}+q_{1}\right)^{2}+\left(\frac{k_{2}}{M}+q_{2}\right)^{2}\right)\tau_{p}^{2}.
\label{eqn:g_gaussian_noise}
\end{eqnarray}    
\end{thm}
\vspace{-2mm}
\begin{IEEEproof}
See Appendix \ref{appendix_b2}.
\end{IEEEproof}
The range of $q_{1}, q_{2}$ in the infinite sum in (\ref{eqn:gaussian_noise_covariance}) can be chosen to be in a reasonable finite range. A range of -20 to 20 for $q_{1}, q_{2}$ in (\ref{eqn:gaussian_noise_covariance}) has been found to be adequate for the computation. Increasing beyond this range has been found to have negligible effect.

\vspace{-2mm}
\section{Closed-form Expressions for Matched Filtering}
\label{sec_mf}
\vspace{-1mm}
In this section, we are interested in exact closed-form expressions for $h_{\mathrm{eff}}[k,l]$ and noise covariance for the case of matched filtering with sinc and Gaussian Tx filters. The exact closed-form expressions for the case of sinc filter have not been reported in the literature, and we derive them here. The exact closed-form expressions for the case of Gaussian filter have been reported in
$\cite{pulse_shaping}$, and we reproduce them here for immediate reference.

The effective channel in the continuous DD domain for the case of matched filtering, i.e., $w_{rx}=w_{tx}^{\dagger}$, can be written as
\vspace{0mm}
\begin{eqnarray}
h_{\mathrm{eff}}(\tau,\nu) & \hspace{-2mm} = & \hspace{-2mm} w_{\mathrm{rx}}(\tau,\nu)*_{\sigma}h_{\mathrm{phy}}(\tau,\nu)*_{\sigma}w_{\mathrm{tx}}(\tau,\nu) \nonumber \\ 
& \hspace{-30mm} = & \hspace{-17mm} w_{\mathrm{rx}}(\tau,\nu)*_{\sigma}\left(\sum_{i=1}^{P}h_{i}\delta(\tau-\tau_{i})\delta(\nu-\nu_{i})\right)*_{\sigma}w_{1}(\tau)w_{2}(\nu) \nonumber    \\ 
& \hspace{-30mm} = & \hspace{-17mm} w_{1}^{*}(-\tau)w_{2}^{*}(-\nu)e^{j2\pi\nu\tau}*\bigg(\sum_{i=1}^{P}h_{i}w_{1}(\tau-\tau_{i})w_{2}(\nu-\nu_{i}) \nonumber \\ 
& \hspace{-30mm} & \hspace{-17mm} e^{j2\pi\nu_{i}(\tau-\tau_{i})}\bigg) \nonumber \\
& \hspace{-30mm} = & \hspace{-17mm} \sum_{i=1}^{P}
\underbrace{\left(\int w_{1}^{*}(-\tau')w_{1}(\tau-\tau_{i}-\tau') e^{-j2\pi\nu_{i}\tau'}d\tau'\right)}_{\overset{\Delta}{=}I_{i}^{(1)}(\tau)} \nonumber \\
& \hspace{-35mm} & \hspace{-20mm} 
\underbrace{\left(\int \hspace{-1mm} w_{2}^{*}(-\nu')w_{2}(\nu-\nu_{i}-\nu')e^{j2\pi\nu'\tau}d\nu'\hspace{-0.5mm} \right)}_{\overset{\Delta}{=}I_{i}^{(2)}(\tau,\nu)}
\hspace{-0.5mm} h_{i}e^{j2\pi\nu_{i}(\tau-\tau_{i})}.
\label{eqn:channel_matched}
\end{eqnarray}
The continuous DD domain noise at the output is
\begin{eqnarray}
n_{\mathrm{dd}}^{w_{\mathrm{rx}}}(\tau,\nu) & \hspace{-2mm} = & \hspace{-2mm} w_{\mathrm{rx}}(\tau,\nu)*_{\sigma}n_{\mathrm{dd}}(\tau,\nu) \nonumber \\ 
& \hspace{-33mm} = & \hspace{-18mm} w_{1}^{*}(-\tau)w_{2}^{*}(-\nu)e^{j2\pi\nu\tau}\hspace{-1mm} *_{\sigma}\hspace{-0.5mm} \Big(\hspace{-0.5mm} \sqrt{\tau_{p}}\sum_{q\in\mathbb{Z}}n(\tau+q\tau_{p})e^{-j2\pi\nu q\tau_{p}}\Big) \nonumber \\ 
& \hspace{-33mm} = & \hspace{-18mm} \sqrt{\tau_{p}}\sum_{q=-\infty}^{\infty}e^{-j2\pi\nu q\tau_{p}}\left(\int w_{1}^{*}(-\tau')n(\tau-\tau'+q\tau_{p})d\tau'\right) \nonumber \\ 
& \hspace{-33mm} & \hspace{-18mm}  \underbrace{\left(\int w_{2}^{*}(-\nu')e^{j2\pi\nu'(\tau+q\tau_{p})}\textcolor{black}{d\nu^{'}}\right)}_{\overset{\Delta}{=}I_{q}^{(3)}(\tau)}.
\label{eqn:noise_matched}
\vspace{-1mm}
\end{eqnarray}
Exact closed-form expressions for $h_{\mathrm{eff}}[k,l]$ and noise covariance for sinc filter are derived in the following subsections.

\vspace{-2mm}
\subsection{$h_{\mathrm{eff}}[k,l]$ for sinc filter}
Integration of the inner integrals defined as $I_{i}^{(1)}(\tau)$ and $I_{i}^{(2)}(\tau,\nu)$, $1\leq i\leq P$, in (\ref{eqn:channel_matched}) gives
\vspace{-2mm}
\begin{eqnarray}
I_{i}^{(1)}(\tau) & \hspace{-2mm} = & \hspace{-2mm}\left(\frac{B-|\nu_{i}|}{B}\right)e^{-j\pi\nu_{i}(\tau-\tau_{i})} \nonumber \\ 
& \hspace{-2mm} & \hspace{-2mm} \mathrm{sinc}((B-|\nu_{i}|)(\tau-\tau_{i}))\mathbbm{1}_{\{-B< \nu_{i}<B\}},
\label{eqn:I1_sinc}
\end{eqnarray}
\vspace{-4mm}
\begin{eqnarray}
I_{i}^{(2)}(\tau, \nu) & \hspace{-2mm} = & \hspace{-2mm} \left(\frac{T-|\tau|}{T}\right)e^{j\pi\tau(\nu-\nu_{i})} \nonumber \\
& \hspace{-2mm} & \hspace{-2mm}\mathrm{sinc}((T-|\tau|)(\nu-\nu_{i}))\mathbbm{1}_{\{-T< \tau< T\}}. 
\label{eqn:I2_sinc}
\end{eqnarray}
Substituting (\ref{eqn:I1_sinc}) and (\ref{eqn:I2_sinc}) in (\ref{eqn:channel_matched}) gives
\vspace{-2mm}
\begin{eqnarray}
\hspace{-4mm}
h_{\mathrm{eff}}(\tau,\nu)& \hspace{-2mm} = & \hspace{-2mm} \sum_{i=1}^{P}h_{i}e^{j\pi(\tau\nu-\tau_{i}\nu_{i})}\left(\frac{T-|\tau|}{T}\right)\left(\frac{B-|\nu_{i}|}{B}\right) \nonumber \\ 
& \hspace{-10mm} & \hspace{-10mm} \mathrm{sinc}((B-|\nu_{i}|)(\tau-\tau_{i})) \mathrm{sinc}((T-|\tau|)(\nu-\nu_{i})) \nonumber \\ 
& \hspace{-10mm} & \hspace{-10mm}\mathbbm{1}_{\{-T<\tau< T\}}\mathbbm{1}_{\{-B<\nu_{i}< B\}}. 
\label{eqn:matched_eff_channel} 
\end{eqnarray}
Sampling (\ref{eqn:matched_eff_channel}) on 
$\Lambda_{\mathrm{dd}}=\{(k\frac{\tau_{p}}{M},l\frac{\nu_{p}}{N})|k,l\in\mathbb{Z}\}$ gives the exact closed-from expression for $h_{\mathrm{eff}}[k,l]$ as
\begin{eqnarray}
h_{\mathrm{eff}}[k,l] & \hspace{-2mm} = & \hspace{-2mm} \sum_{i=1}^{P}h_{i}e^{j\pi\left(\frac{kl}{MN}-\tau_{i}\nu_{i}\right)}\left(\frac{T-|\frac{k\tau_{p}}{M}|}{T}\right)\left(\frac{B-|\nu_{i}|}{B}\right) \nonumber \\ 
& \hspace{-10mm} & \hspace{-10mm} \mathrm{sinc}\left((B-|\nu_{i}|)\left(\frac{k\tau_{p}}{M}-\tau_{i}\right)\right)\mathrm{sinc}\left(\left(T-\left|\frac{k\tau_{p}}{M}\right|\right)\right.
\nonumber \\ 
& \hspace{-10mm} & \hspace{-10mm} 
\left. \left(\frac{l\nu_{p}}{N}-\nu_{i}\right) \right) 
\mathbbm{1}_{\{-T<\frac{k\tau_{p}}{M}< T\}}\mathbbm{1}_{\{-B<\nu_{i}< B\}}.
\label{eqn:MF_IO_sinc}
\end{eqnarray}

\subsection{Noise covariance for sinc filter}
Integration of the inner integral $I_{q}^{(3)}(\tau),q\in\mathbb{Z}$, in (\ref{eqn:noise_matched}) gives
\begin{equation}
I_{q}^{(3)}(\tau) = \frac{1}{\sqrt{T}} \mathrm{rect}\left(\frac{\tau+q\tau_{p}}{T}\right). 
\label{eqn:Iq_noise_matched}
\end{equation}
Substituting (\ref{eqn:Iq_noise_matched}) in (\ref{eqn:noise_matched}) gives
\begin{eqnarray}
n_{\mathrm{dd}}^{w_{\mathrm{rx}}}(\tau,\nu) & \hspace{-2mm} = & \hspace{-2mm} \sqrt{\frac{B\tau_{p}}{T}}\sum_{q=-\infty}^{\infty}e^{-j2\pi\nu q\tau_{p}} \mathrm{rect}\left(\frac{\tau+q\tau_{p}}{T}\right) \nonumber \\ 
& \hspace{-2mm} & \hspace{-2mm} \underbrace{\left(\int \mathrm{sinc}(B\tau')n(\tau-\tau'+q\tau_{p})d\tau'\right)}_{\overset{\Delta}{=}f(\tau+q\tau_{p})}.
\label{eqn:noise_matched_continuous}
\end{eqnarray}
Sampling (\ref{eqn:noise_matched_continuous}) on 
$\Lambda_{\mathrm{dd}}={(k\frac{\tau_{p}}{M},l\frac{\nu_{p}}{N})|k,l\in\mathbb{Z}}$ gives
\begin{eqnarray}
n_{\mathrm{dd}}[k,l] & \hspace{-2mm} = & \hspace{-2mm} \sqrt{\frac{B\tau_{p}}{T}}\sum_{q=-\infty}^{\infty}e^{-j2\pi ql/N} \mathrm{rect}\left(\frac{\frac{k\tau_{p}}{M}+q\tau_{p}}{T}\right) \nonumber \\ 
& \hspace{-2mm} & \hspace{-2mm} f\left(\frac{k\tau_{p}}{M}+q\tau_{p}\right).
\label{eqn:noise_matched_sample}
\end{eqnarray}
From (\ref{eqn:noise_matched_sample}), the expression for the noise covariance can be written as 
\begin{eqnarray}
\hspace{-4mm}
\mathbb{E}[n_{\mathrm{dd}}[k_{1},l_{1}]n_{\mathrm{dd}}^{*}[k_{2},l_{2}]] & \hspace{-2mm} =  & \hspace{-2mm} \left(\frac{B\tau_{p}}{T}\right)\sum_{q_{1}=-\infty}^{\infty}\sum_{q_{2}=-\infty}^{\infty} \nonumber \\ 
& \hspace{-38mm} & \hspace{-38mm} e^{j2\pi\frac{q_{2}l_{2}-q_{1}l_{1}}{N}} \mathrm{rect}\left(\frac{\frac{k_{1}\tau_{p}}{M}+q_{1}\tau_{p}}{T}\right) \mathrm{rect}\left(\frac{\frac{k_{2}\tau_{p}}{M}+q_{2}\tau_{p}}{T}\right) \nonumber \\ 
& \hspace{-38mm} & \hspace{-38mm} \mathbb{E}\bigg[f\left(\frac{k_{1}\tau_{p}}{M}+q_{1}\tau_{p}\right)f^{*}\left(\frac{k_{2}\tau_{p}}{M}+q_{2}\tau_{p}\right)\bigg].
\label{eqn:noise_matched_sample_covariance}
\end{eqnarray}
The term $\mathbb{E}\bigg[f\left(\frac{k_{1}\tau_{p}}{M}+q_{1}\tau_{p}\right)f^{*}\left(\frac{k_{2}\tau_{p}}{M}+q_{2}\tau_{p}\right)\bigg]$ can be solved analytically as 
\begin{eqnarray}
\mathbb{E}\bigg[f\left(\frac{k_{1}\tau_{p}}{M}+q_{1}\tau_{p}\right)f^{*}\left(\frac{k_{2}\tau_{p}}{M}+q_{2}\tau_{p}\right)\bigg] & \hspace{-30mm} & \hspace{-30mm} \nonumber \\
& \hspace{-95mm} = & \hspace{-35mm} \iint \mathrm{sinc}(B\tau_{1}) \ \mathrm{sinc}(B\tau_{2}) \nonumber \\ 
& \hspace{-95mm} & \hspace{-36mm}  \underbrace{\mathbb{E}\bigg[n\left(\frac{k_{1}\tau_{p}}{M}+q_{1}\tau_{p}-\tau_{1}\right)n^{*}\hspace{-1mm} \left(\frac{k_{2}\tau_{p}}{M}+q_{2}\tau_{p}-\tau_{2}\right)\hspace{-1mm}\bigg]}_{=N_{0}\delta\left(\tau_{2}-\tau_{1}-(\frac{k_{2}-k_{1}}{M})\tau_{p}-(q_{2}-q_{1})\tau_{p}\right)}d\tau_{1}d\tau_{2} \nonumber \\ 
& \hspace{-95mm} = & \hspace{-34mm} N_{0}\hspace{-0.5mm}\int \hspace{-0.5mm} \mathrm{sinc}(B\tau_{1})\ \mathrm{sinc}\bigg(B\bigg(\left(\frac{k_{2}-k_{1}}{M}\right)\tau_{p}+(q_{2}-q_{1})\tau_{p} \nonumber \\ 
& \hspace{-95mm} & \hspace{-33mm} + \ \tau_{1}\bigg)\bigg)d\tau_{1} \nonumber \\ 
& \hspace{-95mm} = & \hspace{-34mm} \left(\frac{N_{0}}{B}\right)\mathrm{sinc}\left(B\left(\left(\frac{k_{2}-k_{1}}{M}\right)\tau_{p}+(q_{2}-q_{1})\tau_{p}\right)\right).
\label{eqn:noise_matched_expectation}
\end{eqnarray}
Substituting (\ref{eqn:noise_matched_expectation}) in (\ref{eqn:noise_matched_sample_covariance}) gives the 
closed-form expression
\begin{eqnarray}
\hspace{-4mm}
\mathbb{E}[n_{\mathrm{dd}}[k_{1},l_{1}]n_{\mathrm{dd}}^{*}[k_{2},l_{2}]] & \hspace{-2mm} = & \hspace{-2mm} N_{0}\left(\frac{\tau_{p}}{T}\right)\sum_{q_{1}=-\infty}^{\infty}\sum_{q_{2}=-\infty}^{\infty} \nonumber \\ 
& \hspace{-32mm} & \hspace{-32mm} e^{j2\pi(\frac{q_{2}l_{2}-q_{1}l_{1}}{N})}\mathrm{sinc}
\textcolor{black}{\left(B\left(\left(\frac{k_{2}-k_{1}}{M}\right)\tau_{p}+(q_{2}-q_{1})\tau_{p}\right)\right)} \nonumber \\
& \hspace{-32mm} & \hspace{-32mm}  \mathrm{rect}\left(\frac{\frac{k_{1}\tau_{p}}{M}+q_{1}\tau_{p}}{T}\right) \mathrm{rect}\left(\frac{\frac{k_{2}\tau_{p}}{M}+q_{2}\tau_{p}}{T}\right). 
\label{eqn:MF_covar_sinc}
\end{eqnarray}
Though the range of $q_{1}, q_{2}$ in (\ref{eqn:MF_covar_sinc}) is from $-\infty$ to $+\infty$, the sum will go over only in a finite range due to the presence of the $\mathrm{rect}(.)$ functions.

\vspace{-4mm}
\subsection{Expressions for Gaussian filter}
\label{sec:mf_gauss}
For matched filtering with Gaussian filter, closed-form expressions for $h_{\mathrm{eff}}[k,l]$ 
and $\mathbb{E}[n_{\mathrm{dd}}[k_{1},l_{1}]n_{\mathrm{dd}}^{*}[k_{2},l_{2}]]$  are presented in (34) and (35), respectively, in \cite{pulse_shaping}. We reproduce these equations below for immediate reference:
\begin{eqnarray}
\hspace{-8mm}
h_{\mathrm{eff}}[k,l] & \hspace{-2mm} = & \hspace{-2mm} \sum_{i=1}^{P}h_{i}e^{j\pi\left(\frac{kl}{MN}-\tau_{i}\nu_{i}\right)}e^{-\frac{\alpha_{\tau}B^{2}}{2}\left(\frac{k\tau_{p}}{M}-\tau_{i}\right)^2} \nonumber \\ 
& \hspace{-2mm} & \hspace{-2mm} e^{-\frac{\alpha_{\nu}T^{2}}{2}\left(\frac{l\nu_{p}}{N}-\nu_{i}\right)^2} e^{-\frac{\pi^{2}}{2}\left(\frac{k^{2}\tau_{p}^{2}}{M^{2}\alpha_{\nu}T^{2}}+\frac{\nu_{i}^{2}}{\alpha_{\tau}B^{2}}\right)},
\label{eqn:MF_IO_Gauss}
\end{eqnarray}
\vspace{-4mm}
\begin{eqnarray}
\mathbb{E}[n_{\mathrm{dd}}[k_{1},l_{1}]n_{\mathrm{dd}}^{*}[k_{2},l_{2}]] & \hspace{-2mm} = & \hspace{-2mm} N_{0}\left(\frac{\tau_{p}}{T}\right)\sqrt{\frac{2\pi}{\alpha_{\nu}}} \nonumber \\
& \hspace{-37mm} & \hspace{-37mm} \sum_{q_{1}=-\infty}^{\infty}\sum_{q_{2}=-\infty}^{\infty}e^{j2\pi\frac{q_{2}l_{2}-q_{1}l_{1}}{N}}e^{-\frac{\pi^{2}\tau_{p}^{2}}{\alpha_{\nu}T^{2}}\left(\left(\frac{k_{1}}{M}+q_{1}\right)^{2}+\left(\frac{k_{2}}{M}+q_{2}\right)^{2}\right)} \nonumber \\
& \hspace{-37mm} & \hspace{-37mm} e^{-\frac{\alpha_{\tau}B^{2}}{2}\left(\left(\frac{k_{2}-k_{1}}{M}\right)\tau_{p}+\left(q_{2}-q_{1}\right)\tau_{p}\right)^{2}}.
\label{eqn:MF_covar_Gauss}
\end{eqnarray}

\vspace{-3mm}
\hspace{-1mm}
A range of -20 to 20 for $q_{1}, q_{2}$ has been found to be adequate for accurate computation of   (\ref{eqn:MF_covar_Gauss}).

\section{Closed-form 
Expressions for Channel Matched Filtering}
\label{sec_cmf}
In this section, we derive exact closed-form expressions for $h_{\mathrm{eff}}[k,l]$ and 
$\mathbb{E}[n_{\mathrm{dd}}[k_{1},l_{1}]n_{\mathrm{dd}}^{*}[k_{2},l_{2}]]$ for
channel matched filtering, where the Rx filter is matched to both the Tx filter and the physical channel, i.e., $w_{\mathrm{rx}}=(h_{\mathrm{phy}}(\tau,\nu)*_{\sigma}w_{\tx}(\tau,\nu))^{\dagger}$.
Expressions for both sinc and Gaussian filters are derived.
 
The effective channel in the continuous DD domain for channel matched filtering  
can be written as
\vspace{-1mm}
\begin{eqnarray}
\hspace{-0mm}
h_{\mathrm{eff}}(\tau,\nu) & \hspace{-2mm} = & \hspace{-2mm} w_{\mathrm{rx}}(\tau,\nu)*_{\sigma}h_{\mathrm{phy}}(\tau,\nu)*_{\sigma}w_{\mathrm{tx}}(\tau,\nu) \nonumber \\ 
& \hspace{-32mm} = & \hspace{-17mm} w_{\mathrm{rx}}(\tau,\nu)*_{\sigma}\left(\sum_{i=1}^{P}h_{i}\delta(\tau-\tau_{i})\delta(\nu-\nu_{i})\right)*_{\sigma}w_{1}(\tau)w_{2}(\nu) \nonumber \\
& \hspace{-32mm} = & \hspace{-18mm} \left(\sum_{i=1}^{P}h_{i}^{*}w_{1}^{*}(-\tau-\tau_{i})w_{2}^{*}(-\nu-\nu_{i})e^{j2\pi\nu_{i}(\tau+\tau_{i})}e^{j2\pi\nu\tau}\right)*_{\sigma} \nonumber    \\ 
& \hspace{-32mm} & \hspace{-18mm} \left(\sum_{j=1}^{P}h_{j}w_{1}(\tau-\tau_{j})w_{2}(\nu-\nu_{j})e^{j2\pi\nu_{j}(\tau-\tau_{j})}\right) \nonumber \\ 
& \hspace{-32mm} = & \hspace{-17mm} \sum_{i=1}^{P}\sum_{j=1}^{P}h_{i}^{*}h_{j}e^{j2\pi(\tau_{i}\nu_{i}-\tau_{j}\nu_{j})}e^{j2\pi\tau\nu_{j}} \nonumber \\ 
& \hspace{-32mm} & \hspace{-17mm} \underbrace{\left(\int w_{1}^{*}(-\tau'-\tau_{i})w_{1}(\tau-\tau_{j}-\tau')e^{j2\pi(\nu_{i}-\nu_{j})\tau'}d\tau'\right)}_{\overset{\Delta}{=}I_{ij}^{(1)}(\tau)} \nonumber \\ 
& \hspace{-32mm} & \hspace{-17mm} \underbrace{\left(\int w_{2}^{*}(-\nu'-\nu_{i})w_{2}(\nu-\nu_{j}-\nu')e^{j2\pi\nu'\tau}d\nu'\right)}_{\overset{\Delta}{=}I_{ij}^{(2)}(\tau,\nu)}.
\label{eqn:channel_channel_matched}
\end{eqnarray}
The continuous DD domain noise at the output is
\begin{eqnarray}
n_{\mathrm{dd}}^{w_{\mathrm{rx}}}(\tau,\nu) & \hspace{-2mm} = & \hspace{-2mm} w_{\mathrm{rx}}(\tau,\nu)*_{\sigma}n_{\mathrm{dd}}(\tau,\nu) \nonumber \\ 
& \hspace{-32mm} = & \hspace{-18mm} \left(\sum_{i=1}^{P}\hspace{-0.5mm}h_{i}^{*}w_{1}^{*}(-(\tau+\tau_{i}))w_{2}^{*}(-(\nu+\nu_{i}))e^{j2\pi\nu_{i}(\tau+\tau_{i})+j2\pi\nu\tau}\hspace{-1mm}\right) \nonumber \\ 
& \hspace{-32mm} & \hspace{-18mm} *_{\sigma}\left(\sqrt{\tau_{p}}\sum_{q\in\mathbb{Z}}n(\tau+q\tau_{p})e^{-j2\pi\nu q\tau_{p}}\right) \nonumber \\ 
& \hspace{-32mm} = & \hspace{-18mm} \sqrt{\tau_{p}}\sum_{q=-\infty}^{\infty}\sum_{i=1}^{P}h_{i}^{*}e^{j2\pi\tau_{i}\nu_{i}}e^{-j2\pi\nu q\tau_{p}} \nonumber \\ 
& \hspace{-32mm} & \hspace{-18mm} \left(\int w_{1}^{*}(-\tau'-\tau_{i})n(\tau-\tau'+q\tau_{p})e^{j2\pi\nu_{i}\tau'}d\tau'\right) \nonumber \\ 
& \hspace{-32mm} & \hspace{-18mm} \underbrace{\left(\int w_{2}^{*}(-\nu'-\nu_{i})e^{j2\pi\nu'(\tau+q\tau_{p})}d\nu'\right)}_{\overset{\Delta}{=}I_{iq}^{(3)}(\tau)}.
\label{eqn:noise_channel_matched}
\end{eqnarray}

\subsection{$h_{\mathrm{eff}}[k,l]$ for sinc filter}
Integration of the inner integrals defined as $I_{ij}^{(1)}(\tau)$ and $I_{ij}^{(2)}(\tau,\nu)$, $1\leq i,j\leq P$, in (\ref{eqn:channel_channel_matched}) gives
\begin{eqnarray}
\hspace{-5mm}
I_{ij}^{(1)}(\tau) & \hspace{-2mm} = & \hspace{-2mm} e^{-j2\pi\nu_{ij}\tau_{i}}e^{j\pi\nu_{ij}(\tau+\tau_{ij})}\left(\frac{B-|\nu_{ij}|}{B}\right) \nonumber    \\ 
& \hspace{-2mm} & \hspace{-2mm} \mathrm{sinc}((B-|\nu_{ij}|)(\tau+\tau_{ij}))\mathbbm{1}_{\{-B< \nu_{ij}< B\}}, 
\label{eqn:Iij1_sinc_channel_matched}
\end{eqnarray}
\vspace{-6mm}
\begin{eqnarray}
\hspace{-5mm}
I_{ij}^{(2)}(\tau,\nu) & \hspace{-2mm} = & \hspace{-2mm} e^{-j2\pi\nu_{i}\tau}e^{j\pi\tau(\nu+\nu_{ij})}\left(\frac{T-|\tau|}{T}\right) \nonumber \\ 
& \hspace{-2mm} & \hspace{-2mm} \mathrm{sinc}((T-|\tau|)(\nu+\nu_{ij}))\mathbbm{1}_{\{-T< \tau< T\}},
\label{eqn:Iij2_sinc_channel_matched}
\end{eqnarray}
where $\tau_{ij}=\tau_{i}-\tau_{j}$, $\nu_{ij}=\nu_{i}-\nu_{j}$, $\forall \ 1\leq i,j\leq P$. 
Substituting (\ref{eqn:Iij1_sinc_channel_matched}) and (\ref{eqn:Iij2_sinc_channel_matched}) in (\ref{eqn:channel_channel_matched}) gives
\begin{eqnarray}
\hspace{-3mm}
h_{\mathrm{eff}}(\tau,\nu) & \hspace{-2mm} = & \hspace{-2mm} \sum_{i=1}^{P}\sum_{j=1}^{P}h_{i}^{*}h_{j}e^{j\pi(\tau\nu+\tau_{ij}(\nu_{i}+\nu_{j}))}\left(\frac{B-|\nu_{ij}|}{B}\right) \nonumber \\ 
& \hspace{-18mm} &\hspace{-18mm} \left(\frac{T-|\tau|}{T}\right)\mathrm{sinc}\left(\left(B-|\nu_{ij}|\right)\left(\tau+\tau_{ij}\right)\right)
\nonumber \\ 
& \hspace{-17mm} & \hspace{-17mm} \mathrm{sinc}\left(\left(T-|\tau|\right)\left(\nu+\nu_{ij}\right)\right)\mathbbm{1}_{\{-B<\nu_{ij}< B\}}\mathbbm{1}_{\{-T<\tau< T\}}. 
\label{eqn:sinc_eff_channel_matched}
\end{eqnarray}
Sampling (\ref{eqn:sinc_eff_channel_matched}) on 
$\Lambda_{\mathrm{dd}}=\{(k\frac{\tau_{p}}{M},l\frac{\nu_{p}}{N})|k,l\in\mathbb{Z}\}$ gives the exact closed-form expression for $h_{\mathrm{eff}}[k,l]$ as
\begin{eqnarray}
h_{\mathrm{eff}}[k,l] & \hspace{-2mm} = & \hspace{-2mm} \sum_{i=1}^{P}\sum_{j=1}^{P}h_{i}^{*}h_{j}e^{j\pi\left(\frac{kl}{MN}+\tau_{ij}(\nu_{i}+\nu_{j})\right)}\left(\frac{B-|\nu_{ij}|}{B}\right) \nonumber \\ 
& \hspace{-20mm} & \hspace{-20mm} \left(\frac{T-|\frac{k\tau_{p}}{M}|}{T}\right)\mathrm{sinc}\left(\hspace{-0.5mm}(B-|\nu_{ij}|)\hspace{-0.5mm}\left(\frac{k\tau_{p}}{M}+\tau_{ij}\right)\hspace{-0.5mm}\right) \hspace{-0.5mm}\mathbbm{1}_{\{-B<\nu_{ij}< B\}} \nonumber \\ 
& \hspace{-19mm} & \hspace{-19mm} \mathrm{sinc}\left(\left(T-\bigg|\frac{k\tau_{p}}{M}\bigg|\right)\left(\frac{l\nu_{p}}{N}+\nu_{ij}\right)\right)
\mathbbm{1}_{\{-T<\frac{k\tau_{p}}{M}< T\}}.
\label{eqn:CMF_IO_sinc}
\end{eqnarray}

\vspace{-4mm}
\subsection{Noise covariance for sinc filter}
Integration of the inner integral $I_{iq}^{(3)}(\tau),q\in\mathbb{Z}$, in (\ref{eqn:noise_channel_matched}) gives
\begin{equation}
I_{iq}^{(3)}(\tau) = \frac{1}{\sqrt{T}}e^{-j2\pi\nu_{i}(\tau+q\tau_{p})} \mathrm{rect}\left(\frac{\tau+q\tau_{p}}{T}\right).
\label{eqn:Iiq_sinc_noise_channel_matched}
\end{equation}
Substituting (\ref{eqn:Iiq_sinc_noise_channel_matched}) in (\ref{eqn:noise_channel_matched}) gives
\begin{eqnarray}
\hspace{-5mm}
n_{\mathrm{dd}}^{w_{\mathrm{rx}}}(\tau,\nu) & \hspace{-2mm} = & \hspace{-2mm} \sqrt{\frac{B\tau_{p}}{T}}\sum_{q=-\infty}^{\infty}\sum_{i=1}^{P}h_{i}^{*}e^{j2\pi\tau_{i}\nu_{i}}e^{-j2\pi\nu q\tau_{p}} \nonumber \\ 
& \hspace{-18mm} & \hspace{-18mm} e^{-j2\pi\nu_{i}(\tau+q\tau_{p})} \mathrm{rect}\left(\frac{\tau+q\tau_{p}}{T}\right) \nonumber \\ 
& \hspace{-18mm} & \hspace{-18mm} \underbrace{\left(\int \mathrm{sinc}(B(\tau'+\tau_{i}))n(\tau-\tau'+q\tau_{p})e^{j2\pi\nu_{i}\tau'}d\tau'\right)}_{\overset{\Delta}{=}f_{i}(\tau+q\tau_{p})}.
\label{eqn:sinc_noise_continuous_channel_matched}
\end{eqnarray}
Sampling (\ref{eqn:sinc_noise_continuous_channel_matched}) on 
$\Lambda_{\mathrm{dd}}=\left\{(k\frac{\tau_{p}}{M},l\frac{\nu_{p}}{N})|k,l\in\mathbb{Z}\right\}$ gives
\vspace{-2mm}
\begin{eqnarray}
\hspace{-7mm}
n_{\mathrm{dd}}[k,l] & \hspace{-2mm} = & \hspace{-2mm} \sqrt{\frac{B\tau_{p}}{T}}\sum_{q=-\infty}^{\infty}\sum_{i=1}^{P}h_{i}^{*}e^{j2\pi\tau_{i}\nu_{i}}e^{-j2\pi ql/N} \nonumber \\ 
& \hspace{-17mm} & \hspace{-17mm} e^{-j2\pi\nu_{i}\left(\frac{k\tau_{p}}{M}+q\tau_{p}\right)} \mathrm{rect}\left(\frac{\tau+q\tau_{p}}{T}\right)f_{i}\left(\frac{k\tau_{p}}{M}+q\tau_{p}\right). 
\label{eqn:noise_sample_channel_matched}
\vspace{-2mm}
\end{eqnarray}
From (\ref{eqn:noise_sample_channel_matched}), 
the expression for the noise covariance can be written as
\vspace{-4mm}
\begin{eqnarray}
\mathbb{E}[n_{\mathrm{dd}}[k_{1},l_{1}]n_{\mathrm{dd}}^{*}[k_{2},l_{2}]] & \hspace{-2mm} = & \hspace{-2mm} \left(\frac{B\tau_{p}}{T}\right)\sum_{q_{1}=-\infty}^{\infty}\sum_{q_{2}=-\infty}^{\infty}\sum_{i=1}^{P}\sum_{j=1}^{P} \nonumber \\ 
& \hspace{-42mm} & \hspace{-42mm} h_{i}^{*}h_{j}e^{j2\pi\frac{q_{2}l_{2}-q_{1}l_{1}}{N}}\hspace{-0.5mm} e^{j2\pi(\tau_{i}\nu_{i}-\tau_{j}\nu_{j})}e^{j2\pi\tau_{p}\left(\nu_{j}\left(\frac{k_{2}}{M}+q_{2}\right)-\nu_{i}\left(\frac{k_{1}}{M}+q_{1}\right)\right)} \nonumber \\ 
& \hspace{-42mm} & \hspace{-42mm} \mathbb{E}\bigg[f_{i}\left(\frac{k_{1}\tau_{p}}{M}+q_{1}\tau_{p}\right)f_{j}^{*}\left(\frac{k_{2}\tau_{p}}{M}+q_{2}\tau_{p}\right)\bigg] \nonumber \\ 
& \hspace{-42mm} & \hspace{-42mm}  \mathrm{rect}\left(\frac{\frac{k_{1}\tau_{p}}{M}+q_{1}\tau_{p}}{T}\right) \mathrm{rect}\left(\frac{\frac{k_{2}\tau_{p}}{M}+q_{2}\tau_{p}}{T}\right). 
\label{eqn:sinc_chmatch_noise_covar}
\end{eqnarray}
The term $\mathbb{E}\bigg[f_{i}\left(\frac{k_{1}\tau_{p}}{M}+q_{1}\tau_{p}\right)f_{j}^{*}\left(\frac{k_{2}\tau_{p}}{M}+q_{2}\tau_{p}\right)\bigg]$ in (\ref{eqn:sinc_chmatch_noise_covar}) can be solved analytically as 
\vspace{0mm}
\begin{eqnarray}
\mathbb{E}\bigg[f\left(\frac{k_{1}\tau_{p}}{M}+q_{1}\tau_{p}\right)f^{*}\left(\frac{k_{2}\tau_{p}}{M}+q_{2}\tau_{p}\right)\bigg] & \hspace{-20mm} & \hspace{-20mm} \nonumber \\
& \hspace{-104mm} = & \hspace{-44mm} \iint \mathrm{sinc}(B(\tau_{1}+\tau_{i}))\mathrm{sinc}(B(\tau_{2}+\tau_{j}))e^{j2\pi\nu_{i}\tau_{1}}e^{-j2\pi\nu_{j}\tau_{2}} \nonumber \\
& \hspace{-45mm} & \hspace{-45mm} \underbrace{\mathbb{E}\bigg[n\hspace{-0.5mm}\left(\frac{k_{1}\tau_{p}}{M}+q_{1}\tau_{p}-\tau_{1}\right)n^{*}\hspace{-0.5mm}\left(\frac{k_{2}\tau_{p}}{M}+q_{2}\tau_{p}-\tau_{2}\right)\hspace{-1mm}\bigg]}_{=N_{0}\delta\left(\tau_{2}-\tau_{1}-(\frac{k_{2}-k_{1}}{M})\tau_{p}-(q_{2}-q_{1})\tau_{p}\right)}\hspace{-0.5mm}d\tau_{1}d\tau_{2} \nonumber \\ 
& \hspace{-104mm} = & \hspace{-44mm} N_{0}\hspace{-1.0mm}\int \hspace{-1.0mm} \mathrm{sinc}(B(\tau_{1}+\tau_{i}))\mathrm{sinc}\bigg(\hspace{-0.5mm}B\bigg(\hspace{-0.0mm}\underbrace{\left(\frac{k_{2}-k_{1}}{M}\right)\hspace{-0.5mm}\tau_{p}\hspace{-0.5mm}+\hspace{-0.5mm}(q_{2}-q_{1})\tau_{p}}_{\overset{\Delta}{=}z[k_{1},k_{2},q_{1},q_{2}]} \nonumber \\ 
& \hspace{-45mm} & \hspace{-45mm} +\tau_{1}+\tau_{j}\bigg)\bigg)e^{j2\pi\nu_{i}\tau_{1}}e^{-j2\pi\nu_{j}\left(\left(\frac{k_{2}-k_{1}}{M}\right)\tau_{p}+(q_{2}-q_{1})\tau_{p}+\tau_{1}\right)}d\tau_{1} \nonumber \\ 
& \hspace{-104mm} = & \hspace{-43mm} N_{0}e^{-j\pi(\nu_{i}+\nu_{j})z[k_{1},k_{2},q_{1},q_{2}]}e^{-j\pi(\tau_{i}+\tau_{j})\nu_{ij}}\left(\frac{B-|\nu_{ij}|}{B^{2}}\right) \nonumber \\ 
& \hspace{-45mm} & \hspace{-45mm} \mathrm{sinc}((B-|\nu_{ij}|)(\tau_{ij}-z[k_{1},k_{2},q_{1},q_{2}]))\mathbbm{1}_{\{-B<\nu_{ij}<B\}}.
\label{eqn:sinc_chmatch_expectation}
\end{eqnarray}
Substituting (\ref{eqn:sinc_chmatch_expectation}) in (\ref{eqn:sinc_chmatch_noise_covar}) gives the closed-form expression
\begin{eqnarray}
\mathbb{E}[n_{\mathrm{dd}}[k_{1},l_{1}]n_{\mathrm{dd}}^{*}[k_{2},l_{2}]] & \hspace{-2mm} = & \hspace{-2mm} \left(\frac{N_{0}\tau_{p}}{T}\right)\sum_{q_{1}=-\infty}^{\infty}\sum_{q_{2}=-\infty}^{\infty}\sum_{i=1}^{P}\sum_{j=1}^{P} \nonumber \\ 
& \hspace{-40mm} & \hspace{-40mm} h_{i}^{*}h_{j}\left(\frac{B-|\nu_{ij}|}{B}\right)e^{j2\pi\frac{q_{2}l_{2}-q_{1}l_{1}}{N}} \nonumber \\
& \hspace{-40mm} & \hspace{-40mm}
e^{j2\pi\tau_{p}\left(\nu_{j}\left(\frac{k_{2}}{M}+q_{2}\right)-\nu_{i}\left(\frac{k_{1}}{M}+q_{1}\right)\right)}   
e^{j\pi(\nu_{i}+\nu_{j})(\tau_{ij}-z[k_{1},k_{2},q_{1},q_{2}])} \nonumber \\ 
& \hspace{-40mm} & \hspace{-40mm} 
\mathrm{sinc}((B-|\nu_{ij}|)(\tau_{ij}-z[k_{1},k_{2},q_{1},q_{2}]))\mathbbm{1}_{\{-B<\nu_{ij}<B\}} \nonumber \\ 
& \hspace{-40mm} & \hspace{-40mm}  \mathrm{rect}\left(\frac{\frac{k_{1}\tau_{p}}{M}+q_{1}\tau_{p}}{T}\right) \mathrm{rect}\left(\frac{\frac{k_{2}\tau_{p}}{M}+q_{2}\tau_{p}}{T}\right). 
\label{eqn:CMF_covar_sinc}
\end{eqnarray}
Though the range of $q_{1}, q_{2}$ in (\ref{eqn:CMF_covar_sinc}) is from $-\infty$ to $+\infty$, the sum will go over only in a finite range due to the presence of the $\mathrm{rect}(.)$ functions.

\subsection{$h_{\mathrm{eff}}[k,l]$ for Gaussian filter}
Integration of the inner integrals defined as $I_{ij}^{(1)}(\tau)$ and $I_{ij}^{(2)}(\tau,\nu)$, $1\leq i,j\leq P$, in (\ref{eqn:channel_channel_matched}) gives
\begin{align}
I_{ij}^{(1)}(\tau)=& e^{-\frac{\alpha_{\tau}B^{2}}{2}\left((\tau+\tau_{ij})^{2}-2j(\tau+\tau_{ij})\frac{\pi\nu_{ij}}{\alpha_{\tau}B^{2}}+\left(\frac{\pi\nu_{ij}}{\alpha_{\tau}B^{2}}\right)^{2}\right)} \nonumber \\
& e^{-j2\pi\nu_{ij}\tau_{i}},
\label{eqn:Iij1_gauss_channel_matched}
\end{align}
\begin{align}
I_{ij}^{(2)}(\tau,\nu)=& e^{-\frac{\alpha_{\nu}T^{2}}{2}\left((\nu+\nu_{ij})^{2}-2j(\nu+\nu_{ij})\frac{\pi\tau}{\alpha_{\nu}T^{2}}+\left(\frac{\pi\tau}{\alpha_{\nu}T^{2}}\right)^{2}\right)} \nonumber \\
&e^{-j2\pi\nu_{i}\tau}.
\label{eqn:Iij2_gauss_channel_matched}
\end{align}
Substituting (\ref{eqn:Iij1_gauss_channel_matched}) and (\ref{eqn:Iij2_gauss_channel_matched}) in (\ref{eqn:channel_channel_matched}) gives
\begin{eqnarray}
h_{\mathrm{eff}}(\tau,\nu) & \hspace{-2mm} = & \hspace{-2mm} \sum_{i=1}^{P}\sum_{j=1}^{P}h_{i}^{*}h_{j}e^{j\pi(\tau\nu+\tau_{ij}(\nu_{i}+\nu_{j}))}e^{-\frac{\alpha_{\tau}B^{2}}{2}(\tau+\tau_{ij})^{2}} \nonumber \\ 
& \hspace{-2mm} & \hspace{-2mm} e^{-\frac{\alpha_{\nu}T^{2}}{2}(\nu+\nu_{ij})^{2}}e^{-\frac{\pi^{2}}{2}\left(\frac{\tau^{2}}{\alpha_{\nu}T^{2}}+\frac{\nu_{ij}^{2}}{\alpha_{\tau}B^{2}}\right)}.
\label{eqn:gauss_eff_channel_matched}
\end{eqnarray}
Sampling (\ref{eqn:gauss_eff_channel_matched}) on 
$\Lambda_{\mathrm{dd}}=\{(k\frac{\tau_{p}}{M},l\frac{\nu_{p}}{N}) | k,l\in \mathbb{Z}\}$ gives
\begin{eqnarray}
h_{\mathrm{eff}}[k,l] & \hspace{-3mm} = & \hspace{-3mm} \sum_{i=1}^{P}\sum_{j=1}^{P}h_{i}^{*}h_{j}e^{j\pi\left(\frac{kl}{MN}+\tau_{ij}(\nu_{i}+\nu_{j})\right)}e^{-\frac{\alpha_{\tau}B^{2}}{2}\left(\frac{k\tau_{p}}{M}+\tau_{ij}\right)^{2}} \nonumber \\ 
& \hspace{-3mm} & \hspace{-3mm} e^{-\frac{\alpha_{\nu}T^{2}}{2}\left(\frac{l\nu_{p}}{N}+\nu_{ij}\right)^{2}}e^{-\frac{\pi^{2}}{2}\left(\frac{k^{2}\tau_{p}^{2}}{M^{2}\alpha_{\nu}T^{2}}+\frac{\nu_{ij}^{2}}{\alpha_{\tau}B^{2}}\right)}.
\label{eqn:gauss_eff_channel_matched_samples}
\end{eqnarray}

\subsection{Noise covariance for Gaussian filter}
Integration of the inner integral $I_{iq}^{(3)}(\tau), q\in\mathbb{Z}$, in (\ref{eqn:noise_channel_matched}) gives
\begin{equation}
I_{iq}^{(3)}(\tau)=\left(\frac{2\pi}{\alpha_{\nu}T^{2}}\right)^{\frac{1}{4}}e^{-j2\pi\nu_{i}(\tau+q\tau_{p})}e^{-\frac{\pi^{2}(\tau+q\tau_{p})^{2}}{\alpha_{\nu}T^{2}}}.
\label{eqn:Iiq_gauss_noise_channel_matched}
\end{equation}
Substituting (\ref{eqn:Iiq_gauss_noise_channel_matched}) in (\ref{eqn:noise_channel_matched}) gives
\begin{eqnarray}
\hspace{-6mm}
n_{\mathrm{dd}}^{w_{\mathrm{rx}}}(\tau,\nu) & \hspace{-2mm} = & \hspace{-2mm} \sqrt{\frac{2B\tau_{p}}{T}}\left(\frac{\alpha_{\tau}}{\alpha_{\nu}}\right)^{\frac{1}{4}}\sum_{q=-\infty}^{\infty}\sum_{i=1}^{P}h_{i}^{*}e^{j2\pi\tau_{i}\nu_{i}} \nonumber    \\ 
& \hspace{-2mm} & \hspace{-2mm} e^{-j2\pi\nu q\tau_{p}}e^{-j2\pi\nu_{i}(\tau+q\tau_{p})}e^{-\frac{\pi^{2}(\tau+q\tau_{p})^{2}}{\alpha_{\nu}T^{2}}} \nonumber    \\ 
& \hspace{-2mm} & \hspace{-2mm} \underbrace{\left(\int \hspace{-1mm} e^{-\alpha_{\tau}B^{2}(\tau_{1}+\tau_{i})^{2}}n(\tau-\tau_{1}+q\tau_{p})d\tau_{1}\right)}_{\overset{\Delta}{=}f_{i}(\tau+q\tau_{p})}\hspace{-0.5mm}.
\label{eqn:gauss_noise_continuous_channel_matched}
\end{eqnarray}
Sampling (\ref{eqn:gauss_noise_continuous_channel_matched}) on 
$\Lambda_{\mathrm{dd}}=\left\{(k\frac{\tau_{p}}{M},l\frac{\nu_{p}}{N})|k,l\in\mathbb{Z}\right\}$ gives
\begin{eqnarray}
n_{\mathrm{dd}}[k,l]& \hspace{-2mm} = & \hspace{-2mm} \sqrt{\frac{2B\tau_{p}}{T}}\left(\frac{\alpha_{\tau}}{\alpha_{\nu}}\right)^{\frac{1}{4}}\sum_{q=-\infty}^{\infty}\sum_{i=1}^{P}h_{i}^{*}e^{j2\pi\tau_{i}\nu_{i}}e^{-j2\pi\frac{ql}{N}} \nonumber \\ 
& \hspace{-10mm} & \hspace{-10mm} e^{-\frac{\pi^{2}\tau_{p}^{2}\left(\frac{k}{M}+q\right)^{2}}{\alpha_{\nu}T^{2}}}e^{-j2\pi\nu_{i}\tau_{p}\left(\frac{k}{M}+q\right)}f_{i}\left(\frac{k\tau_{p}}{M}+q\tau_{p}\right).
\label{eqn:gauss_noise_channel_matched_samples}
\end{eqnarray}
From (\ref{eqn:gauss_noise_channel_matched_samples}), we can write
\begin{eqnarray}
\mathbb{E}[n_{\mathrm{dd}}[k_{1},l_{1}]n_{\mathrm{dd}}^{*}[k_{2},l_{2}]] & \hspace{-2mm} = & \hspace{-2mm} \left(\frac{2B\tau_{p}}{T}\right)\sqrt{\frac{\alpha_{\tau}}{\alpha_{\nu}}}\sum_{q_{1}=-\infty}^{\infty}\sum_{q_{2}=-\infty}^{\infty} \nonumber \\ 
& \hspace{-40mm} & \hspace{-40mm}  \sum_{i=1}^{P}\sum_{j=1}^{P}h_{i}^{*}h_{j}e^{j2\pi\frac{q_{2}l_{2}-q_{1}l_{1}}{N}}e^{-\frac{\pi^{2}\tau_{p}^{2}}{\alpha_{\nu}T^{2}}\left(\left(\frac{k_{1}}{M}+q_{1}\right)^{2}+\left(\frac{k_{2}}{M}+q_{2}\right)^{2}\right)} \nonumber \\ 
& \hspace{-40mm} & \hspace{-40mm} e^{j2\pi(\tau_{i}\nu_{i}-\tau_{j}\nu_{j})}e^{j2\pi\tau_{p}\big[\nu_{j}\left(\frac{k_{2}}{M}+q_{2}\right)-\nu_{i}\left(\frac{k_{1}}{M}+q_{1}\right)\big]} \nonumber \\
& \hspace{-40mm} & \hspace{-40mm} \mathbb{E}\bigg[f_{i}\left(\frac{k_{1}\tau_{p}}{M}+q_{1}\tau_{p}\right)f_{j}^{*}\left(\frac{k_{2}\tau_{p}}{M}+q_{2}\tau_{p}\right)\bigg].
\label{eqn:gauss_channel_matched_samples_covariance}
\end{eqnarray}
The term $\mathbb{E}\bigg[f_{i}\left(\frac{k_{1}\tau_{p}}{M}+q_{1}\tau_{p}\right)f_{j}^{*}\left(\frac{k_{2}\tau_{p}}{M}+q_{2}\tau_{p}\right)\bigg]$ in (\ref{eqn:gauss_channel_matched_samples_covariance}) can be solved analytically as 
\begin{eqnarray}
& \hspace{-8mm} & \hspace{-8mm}
\mathbb{E}\bigg[f_{i}\left(\frac{k_{1}\tau_{p}}{M}+q_{1}\tau_{p}\right)f_{j}^{*}\left(\frac{k_{2}\tau_{p}}{M}+q_{2}\tau_{p}\right)\bigg]  \nonumber \\ 
& \hspace{-6mm} = & \hspace{-4mm} \iint e^{-\alpha_{\tau}B^{2}(\tau_{1}+\tau_{i})^{2}}e^{-\alpha_{\tau}B^{2}(\tau_{2}+\tau_{j})^{2}} \nonumber  \\ 
& \hspace{-8mm} & \hspace{-8mm} \underbrace{\mathbb{E}\bigg[n\left(\frac{k_{1}\tau_{p}}{M}-\tau_{1}+q_{1}\tau_{p}\right)n^{*}\hspace{-1mm}\left(\frac{k_{2}\tau_{p}}{M}-\tau_{2}+q_{2}\tau_{p}\right)\bigg]}_{\overset{\Delta}{=}N_{0}\delta\left(\tau_{2}-\tau_{1}-\left(\frac{k_{2}-k_{1}}{M}\right)\tau_{p}-(q_{2}-q_{1})\tau_{p}\right)}d\tau_{1}d\tau_{2} \nonumber \\ 
& \hspace{-6mm} = & \hspace{-3mm} N_{0}\int e^{-\alpha_{\tau}B^{2}(\tau_{1}+\tau_{i})^{2}}e^{-\alpha_{\tau}B^{2}(\tau_{1}+z[k_{1},k_{2},q_{1},q_{2}]+\tau_{j})^{2}}d\tau_{1} \nonumber  \\ 
& \hspace{-6mm} = & \hspace{-3mm} N_{0}\sqrt{\frac{\pi}{2\alpha_{\tau}B^{2}}}e^{-\frac{\alpha_{\tau}B^{2}}{2}(\tau_{ij}-z[k_{1},k_{2},q_{1},q_{2}])^{2}}.
\label{eqn:expectation_gauss_noise_channel_matched}
\end{eqnarray}
where $z[k_{1},k_{2},q_{1},q_{2}]$ in (\ref{eqn:expectation_gauss_noise_channel_matched}) is same as defined in (\ref{eqn:sinc_chmatch_expectation}). Substituting (\ref{eqn:expectation_gauss_noise_channel_matched}) in (\ref{eqn:gauss_channel_matched_samples_covariance}) gives
\begin{eqnarray}
\mathbb{E}[n_{\mathrm{dd}}[k_{1},l_{1}]n_{\mathrm{dd}}^{*}[k_{2},l_{2}]] & \hspace{-2mm} = & \hspace{-2mm} 
N_{0}\left(\frac{\tau_{p}}{T}\right)\sqrt{\frac{2\pi}{\alpha_{\nu}}}\sum_{q_{1},q_{2}=-\infty}^{\infty}\sum_{i=1}^{P}\sum_{j=1}^{P}    \nonumber    \\ 
& \hspace{-41mm} & \hspace{-41mm} 
h_{i}^{*}h_{j}e^{j2\pi\frac{q_{2}l_{2}-q_{1}l_{1}}{N}}e^{-\frac{\pi^{2}\tau_{p}^{2}}{\alpha_{\nu}T^{2}}\left(\left(\frac{k_{1}}{M}+q_{1}\right)^{2}+\left(\frac{k_{2}}{M}+q_{2}\right)^{2}\right)}e^{j2\pi(\tau_{i}\nu_{i}-\tau_{j}\nu_{j})} \nonumber \\
& \hspace{-41mm} & \hspace{-41mm} 
e^{j2\pi\tau_{p}\big[\nu_{j}\left(\frac{k_{2}}{M}+q_{2}\right)-\nu_{i}\left(\frac{k_{1}}{M}+q_{1}\right)\big]}e^{-\frac{\alpha_{\tau}B^{2}}{2}(\tau_{ij}-z[k_{1},k_{2},q_{1},q_{2}])^{2}}.
\label{eqn:CMF_covar_Gauss}
\end{eqnarray}
A range of -20 to 20 for $q_{1}, q_{2}$ has been found to be adequate for accurate computation of (\ref{eqn:CMF_covar_Gauss}). 

\section{Results and Discussions}
\label{sec4}
In this section, we present the numerical results on the bit error rate (BER) performance of Zak-OTFS for different Tx/Rx DD filter configurations. The BER performance of identical filtering, matched filtering, and channel matched filtering using sinc and Gaussian filters are evaluated and compared. A Zak-OTFS system with $M=12$ and $N=14$ is considered. The Doppler period is fixed at $\nu_{\mathrm p}=15$ kHz and the delay period, therefore, is $\tau_{\mathrm p}=\frac{1}{\nu_{\mathrm p}}=66.66\ \mu$s. Consequently, the time duration of a Zak-OTFS frame is $T=N\tau_{\mathrm p}=0.93$ ms and the bandwidth is $B=M\nu_{\mathrm p}=180$ kHz. We consider the 
Veh-A channel model \cite{vehA} having $P=6$ 
paths with fractional DDs and a power delay profile (PDP) as detailed in Table \ref{tab_pdp}. 
\begin{table}[h]
\centering
\begin{tabular}{|c|c|c|c|c|c|c|}
\hline
Path index ($i$)         & 1 & 2    & 3    & 4    & 5    & 6    \\ \hline
Delay $\tau_{i}$ ($\mu s$)      & 0 & 0.31 & 0.71 & 1.09 & 1.73 & 2.51 \\ \hline
Relative power (dB) & 0 & -1   & -9   & -10  & -15  & -20  \\ \hline
\end{tabular}
\caption{Power delay profile of Veh-A channel model.}
\label{tab_pdp}
\vspace{-2mm}
\end{table}
The Doppler shift of the $i$th path is modeled as $\nu_{i}=\nu_{\mathrm{max}}\cos\theta_{i},i=1,\ldots,P$, where $\theta_{i}$s are independent and uniformly distributed in $[0,2\pi)$. Perfect knowledge of the I/O relation is assumed at the receiver. Also, in the simulations, the range of values of $m$ and $n$ in (\ref{eqn_channel_matrix}) is limited to -2 to 2, and this is found to ensure an adequate support set of $h_{\mathrm{eff}}[k,l]$ that captures the channel spread accurately.
\begin{figure}
\centering
\includegraphics[width=9.5cm,height=7.0cm]{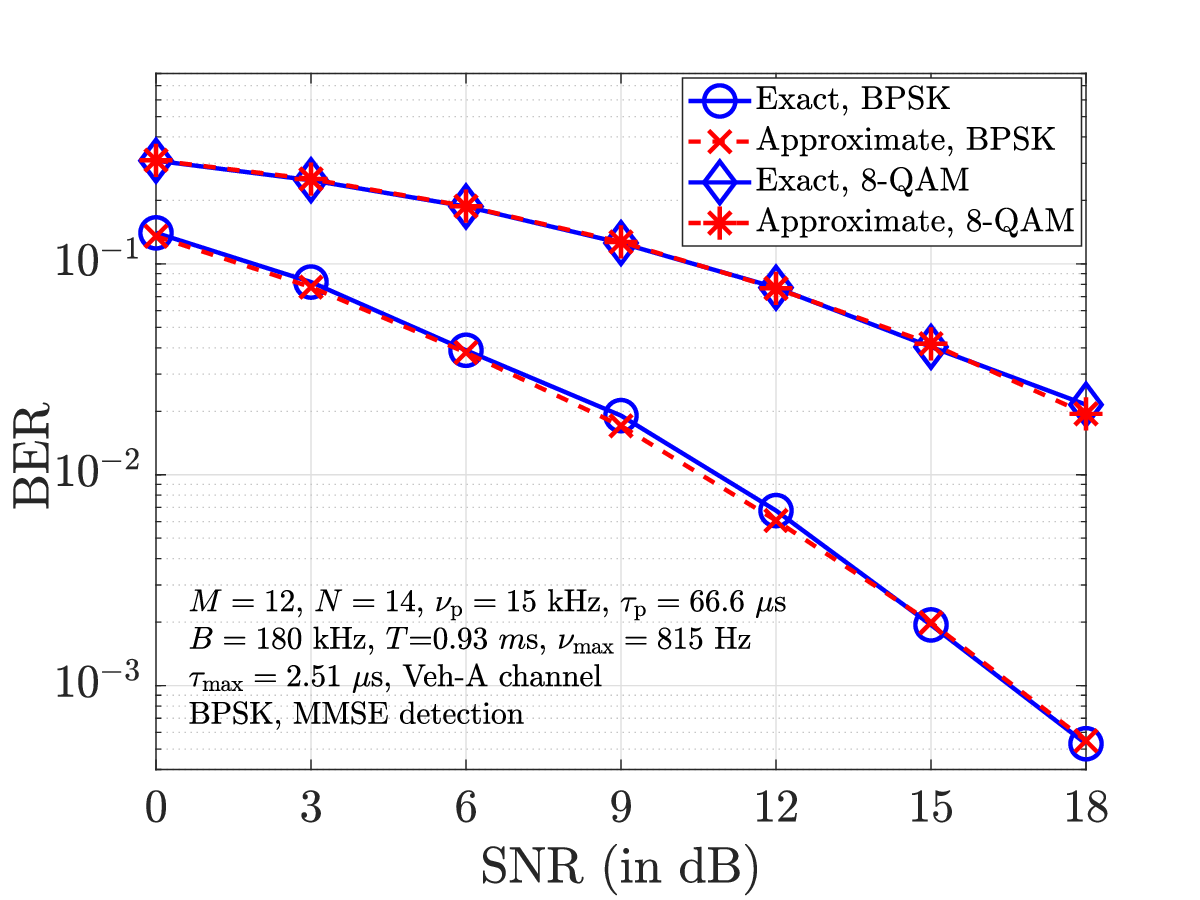}
\caption{BER performance of Zak-OTFS for 
identical filtering with sinc filter using approximate closed-form and exact I/O relation expressions.}
\label{fig:sinc_identical_filter}
\vspace{-5mm}
\end{figure}

For identical filtering with sinc filter, we have obtained approximate closed-form expressions for $h_{\mathrm{eff}}[k,l]$ and noise covariance in 
(\ref{eqn:sinc_identical_closed}) and (\ref{eqn:noise_approx_closed}), whose exact expressions are obtained by computing (\ref{eqn:Exact_identical_channel}) and (\ref{eqn:noise_covariance_identical}), respectively, and sampling on $\Lambda_{\mathrm{dd}}$. In Fig. \ref{fig:sinc_identical_filter}, we assess the accuracy of the approximation by comparing the BER performance obtained using the exact and approximate expressions. BPSK and 8-QAM modulation, MMSE detection, and Veh-A channel model with $\nu_{\mathrm{max}}=815$ Hz are considered. At a carrier frequency of $f_c=4$ GHz and speed of light $c=3\times 10^{8}$ m/s, the maximum Doppler $\nu_\text{max}=815$ Hz corresponds to a mobile speed of $v=\frac{c\nu_\text{max}}{f_c}=$ 61.125 m/s = 220 km/h. From Fig. \ref{fig:sinc_identical_filter}, we observe that the BER performance evaluated using the approximate closed-form expressions is almost same as that evaluated using the exact expressions, demonstrating the accuracy of the approximation. 
\begin{figure}
\centering
\includegraphics[width=9.5cm,height=7.0cm]{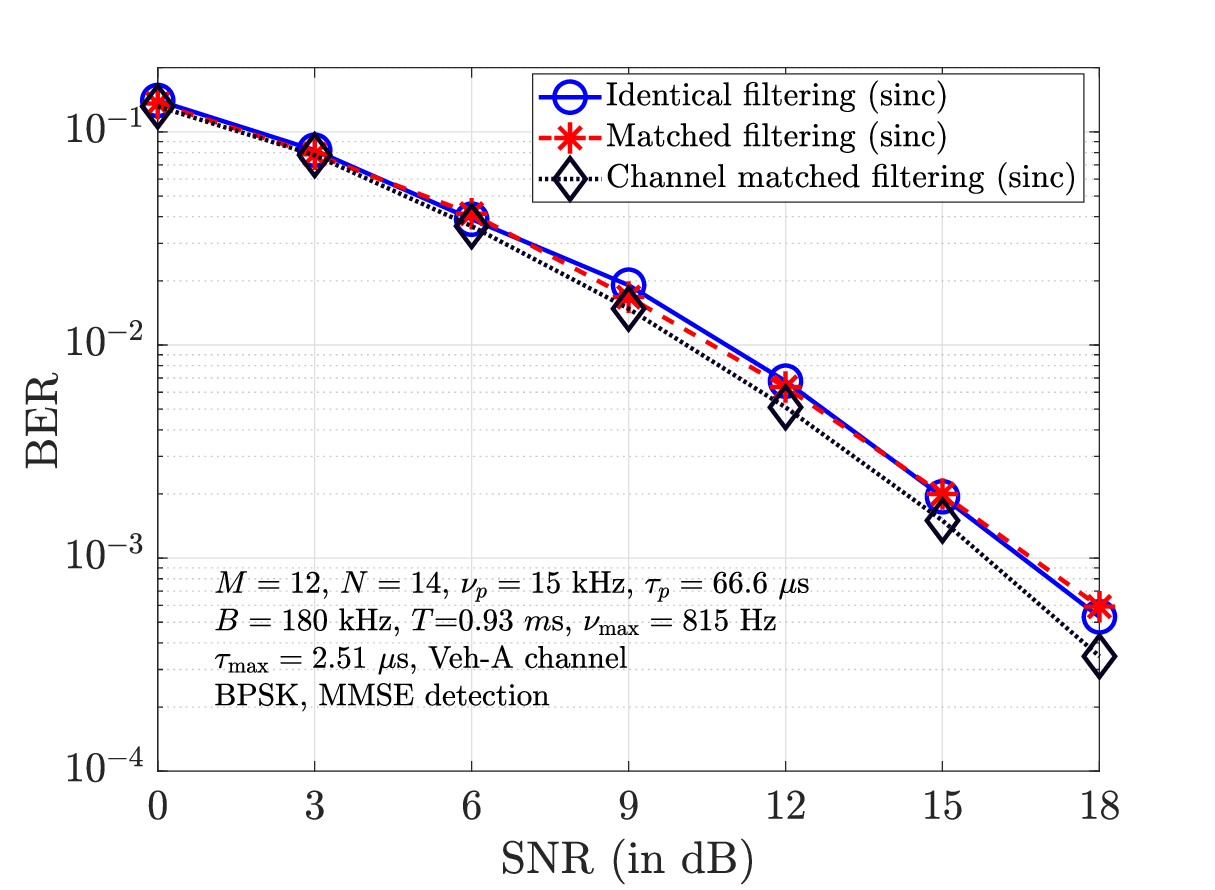}
\caption{BER performance of Zak-OTFS for identical filtering, matched filtering, and channel matched filtering with sinc filter.}
\label{fig:sinc_all_filter}
\vspace{-6mm}
\end{figure}

Next, in Fig. \ref{fig:sinc_all_filter}, we compare the BER performance of Zak-OTFS for identical filtering, matched filtering, and channel matched filtering with sinc filter, BPSK, MMSE detection, and
Veh-A channel model with $\nu_{\mathrm{max}}=815$ Hz.
From Fig. \ref{fig:sinc_all_filter}, we observe that the BER performance for identical filtering and matched filtering are nearly the same, and that the performance of channel matched filtering is the best among the three (better by about 1 dB at $10^{-3}$ BER). This is in line with the result in \cite{ref6} which shows that channel matched filtering operation at the receiver maximizes the SNR. Further, we note that the derived closed-form expressions can avoid the computationally intensive numerical evaluation of the integrals involved in the computation of the effective channel taps $h_\text{eff}[k,l]$s and the noise covariance matrix, yielding simulation speedups. While the computation of $h_\text{eff}[k,l]$s  needs to be done for each channel realization, the noise covariance matrix computation needs to be done only once. The simulation times taken to compute the noise covariance matrix with closed-form expression and without closed-form expression (i.e., with numerical integration) are found to be 0.23 s and 26.05 min, respectively\footnote{The simulations are run on a PC with a 13th Gen Intel Core i7-13700 processor (16 cores, 24 threads) and 48.0 GiB of RAM, using MATLAB R2023b.}, for the system parameters in Fig. \ref{fig:sinc_all_filter} with sinc filter and matched filtering at the receiver. Also, the corresponding simulation times taken for the computation of $h_\text{eff}[k,l]$s for 1000 channel realizations at a given SNR are found to be 0.42 s and 92.01 min, respectively. This, in turn, speeds up the BER simulations, e.g., 1.544 min and 118.77 min simulation times for the cases with closed-form expression and with numerical integration, respectively, for 1000 channel realizations at a given SNR.
\begin{figure}
\centering
\includegraphics[width=9.5cm,height=7.0cm]{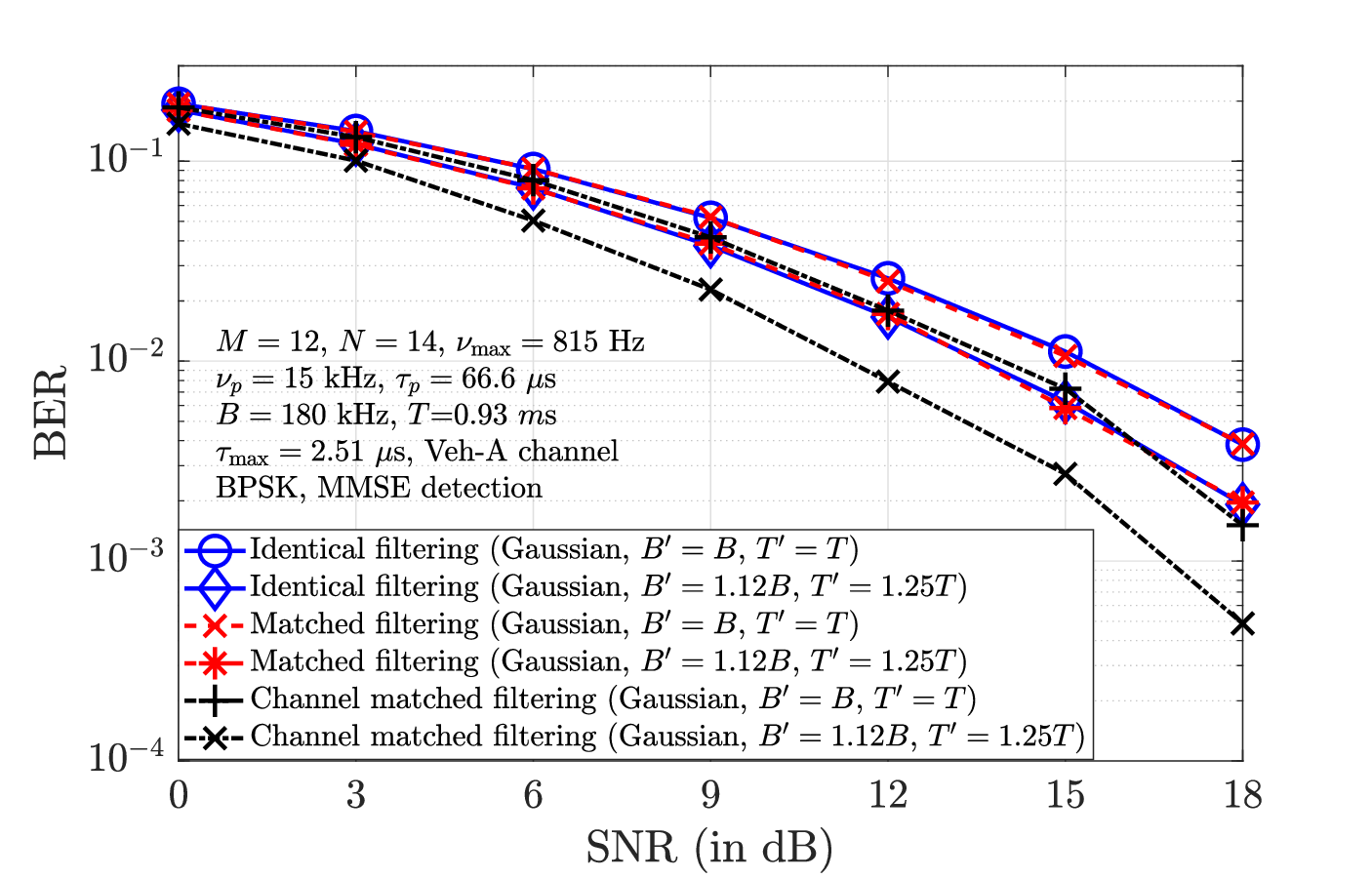}
\caption{BER performance of Zak-OTFS for identical filtering, matched filtering, and channel matched filtering with Gaussian filter.}
\label{fig:Gauss_all_filter}
\vspace{-6mm}
\end{figure}

In Fig. \ref{fig:Gauss_all_filter}, we present a comparison of the BER performance of identical filtering, matched filtering, and channel matched filtering when Gaussian filter is used, with BPSK, MMSE detection, and Veh-A channel model with $\nu_{\mathrm{max}}=815$ Hz. We examine two types of Gaussian filters, namely, $(i)$ Gaussian filter without bandwidth and time expansion (i.e., $B'=B$ and $T'=T$), and $(ii)$ Gaussian filter with bandwidth and time expansion where $B'=1.12B$ and $T'=1.25T$. The following observations can be made from Fig. \ref{fig:Gauss_all_filter}. As observed in the case of sinc filter in Fig. \ref{fig:sinc_all_filter}, the performance of identical filtering and matched filtering with Gaussian filter are almost the same, and the performance of channel matched filtering is better than those of identical filtering and matched filtering. Also, with expansion in bandwidth and time of the Gaussian filter ($B'=1.12B$ and $T'=1.25T$), the performance improves by about 2 dB at high SNRs compared to the performance without bandwidth and time expansion ($B'=B$, $T'=T$). This improvement in BER is achieved at the cost of increased bandwidth and time. 
\vspace{0mm}
\begin{figure}
\centering
\subfloat[Sinc filter]  {\includegraphics[width=4.35cm,height=4.5cm]{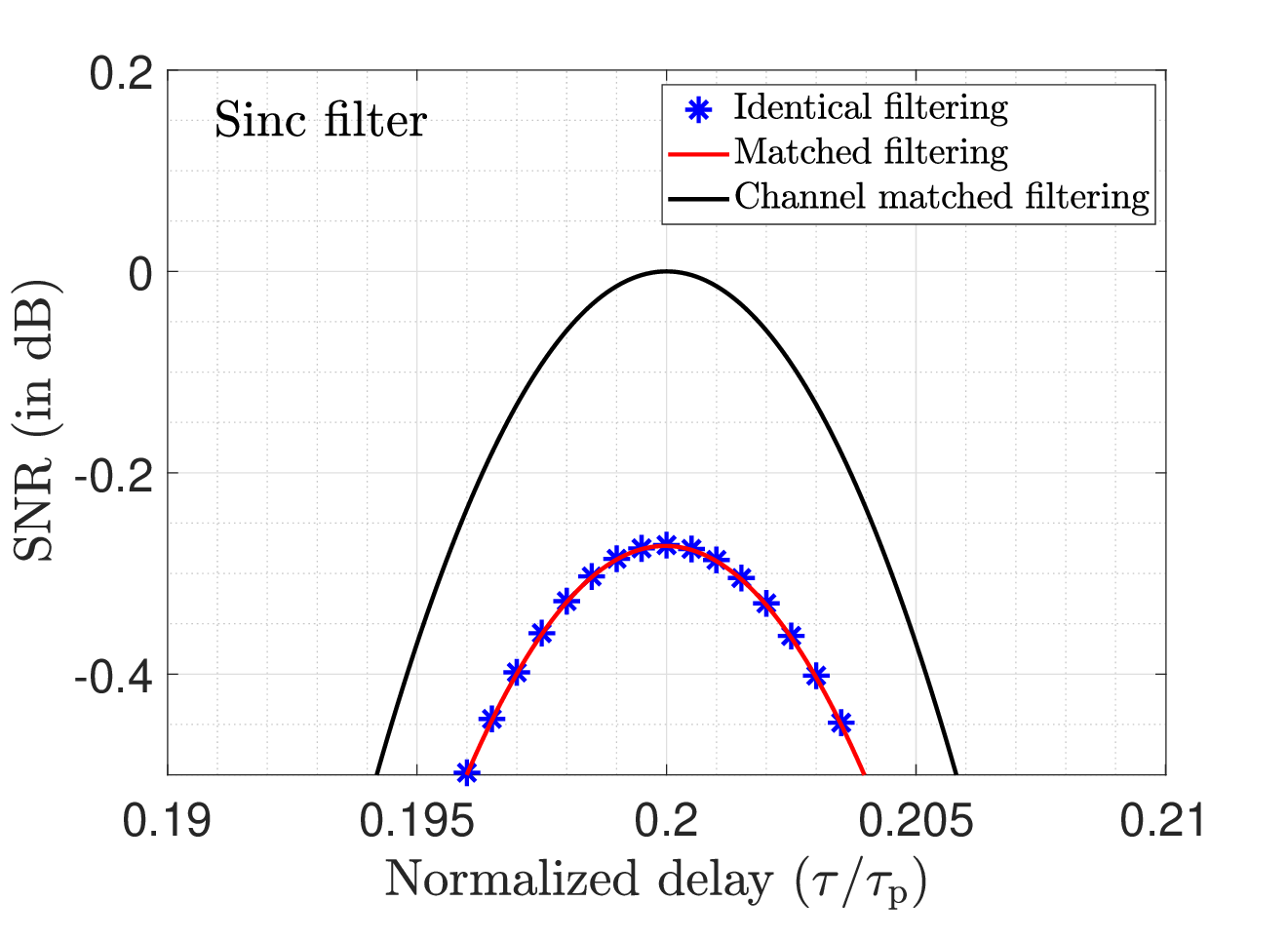} \label{snr_sinc_a}
}  
\subfloat[Gaussian filter]
{\includegraphics[width=4.35cm, height=4.5cm]{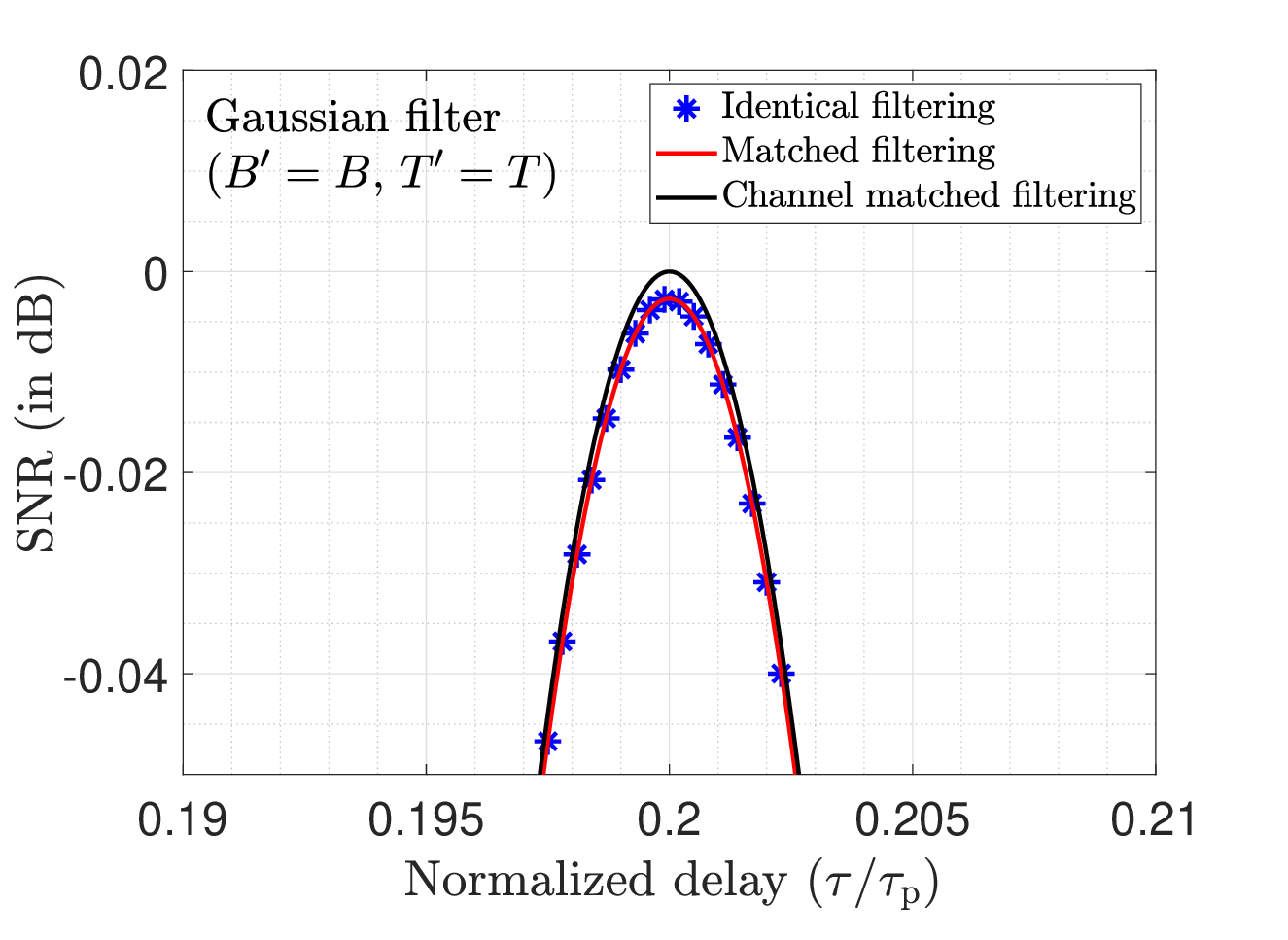} \label{snr_gauss_b}
}
\caption{SNR performance of identical, matched, and channel matched filtering as a function of normalized delay ($\tau/\tau_\text{p}$) at $\nu=\nu_1$ for (a) sinc filter and (b) for Gaussian filter.} 
\label{snr}
\vspace{-4mm}
\end{figure}

{\em Remark:} As mentioned earlier, it has been shown in \cite{ref6} (Theorem 5) that, for any arbitrary Tx filter $w_\text{tx}(\tau,\nu)$, the optimal Rx filter that maximizes the SNR is the one that is matched to the physical channel $h_\text{phy}(\tau,\nu)$ and the Tx filter $w_\text{tx}(\tau,\nu)$, which we have termed as the channel matched filtering in this paper. Here, in Figs. \ref{snr_sinc_a} and \ref{snr_gauss_b}, we present an SNR performance comparison between the three filtering schemes, for sinc and Gaussian filters. For illustration purposes, these SNR plots are generated for a system with $M=N=32$ using a frame consisting of a symbol +1 at location (0,0) and zeros elsewhere, assuming a channel with $P=1$ and parameters $(h_1,\tau_1,\nu_1) =(1,0.2\tau_\text{p},-0.25\nu_\text{p})$. The SNRs are plotted as a function of normalized delay 
$\tau/\tau_\text{p}$ at $\nu=\nu_1$.
It can be seen that while identical filtering and matched filtering achieve almost the same SNR performance, channel matched filtering achieves the highest maximum SNR. This, in turn, leads to the better BER performance of channel matched filtering as observed in Figs. \ref{fig:sinc_all_filter} and \ref{fig:Gauss_all_filter}. 

\begin{figure}
\centering
\includegraphics[width=9.5cm,height=7cm]{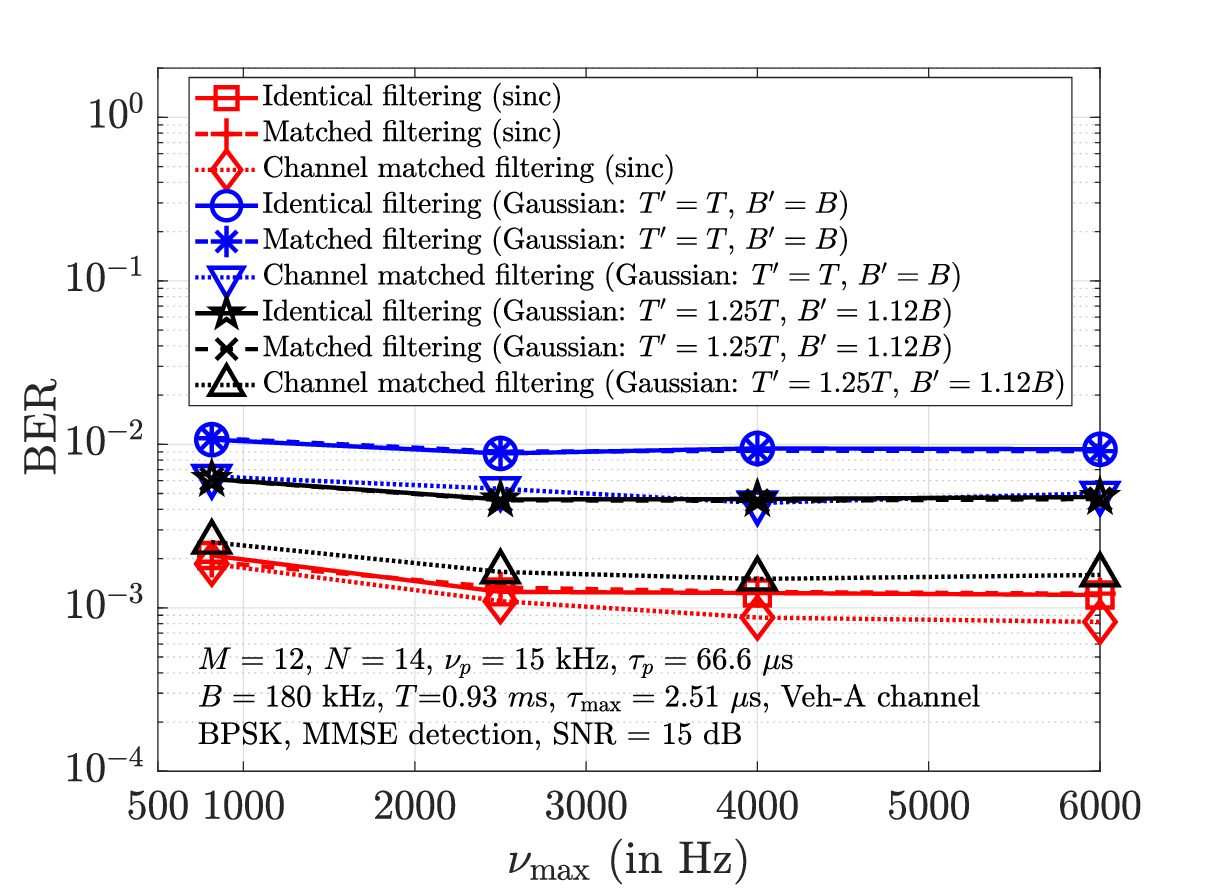}
\caption{BER versus $\nu_{\max}$ performance of Zak-OTFS with identical filtering, matched filtering, and channel matched filtering with sinc and Gaussian filters.}
\label{fig:numax}
\vspace{-3mm}
\end{figure}

Next, in Fig. \ref{fig:numax}, we present the BER performance as a function of $\nu_{\textrm{max}}$ at SNR = 15 dB with BPSK, MMSE detection, and Veh-A channel model for identical filtering, matched filtering, and channel matched filtering with sinc and Gaussian filters. First, the robustness of Zak-OTFS for increased Dopplers can be observed in this figure. It can also be observed that sinc filter achieves better performance compared to Gaussian filter. This can be attributed to the fact that the sinc filter has nulls at points on $\Lambda_{\mathrm{dd}}$ (whereas Gaussian filter has non-zero values at these points), which,
in the presence of perfect knowledge of the I/O relation, leads to the better performance of sinc filter. Further, channel matched filtering is observed to achieve better performance when sinc filter is used compared to when Gaussian filter is used even with some bandwidth/time expansion 
$(B'\hspace{-0mm}=\hspace{-0mm}1.12B$, $T'\hspace{-0mm}=\hspace{-0mm}1.25T)$. 

\begin{figure}
\centering
\includegraphics[width=9.5cm,height=6.5cm]{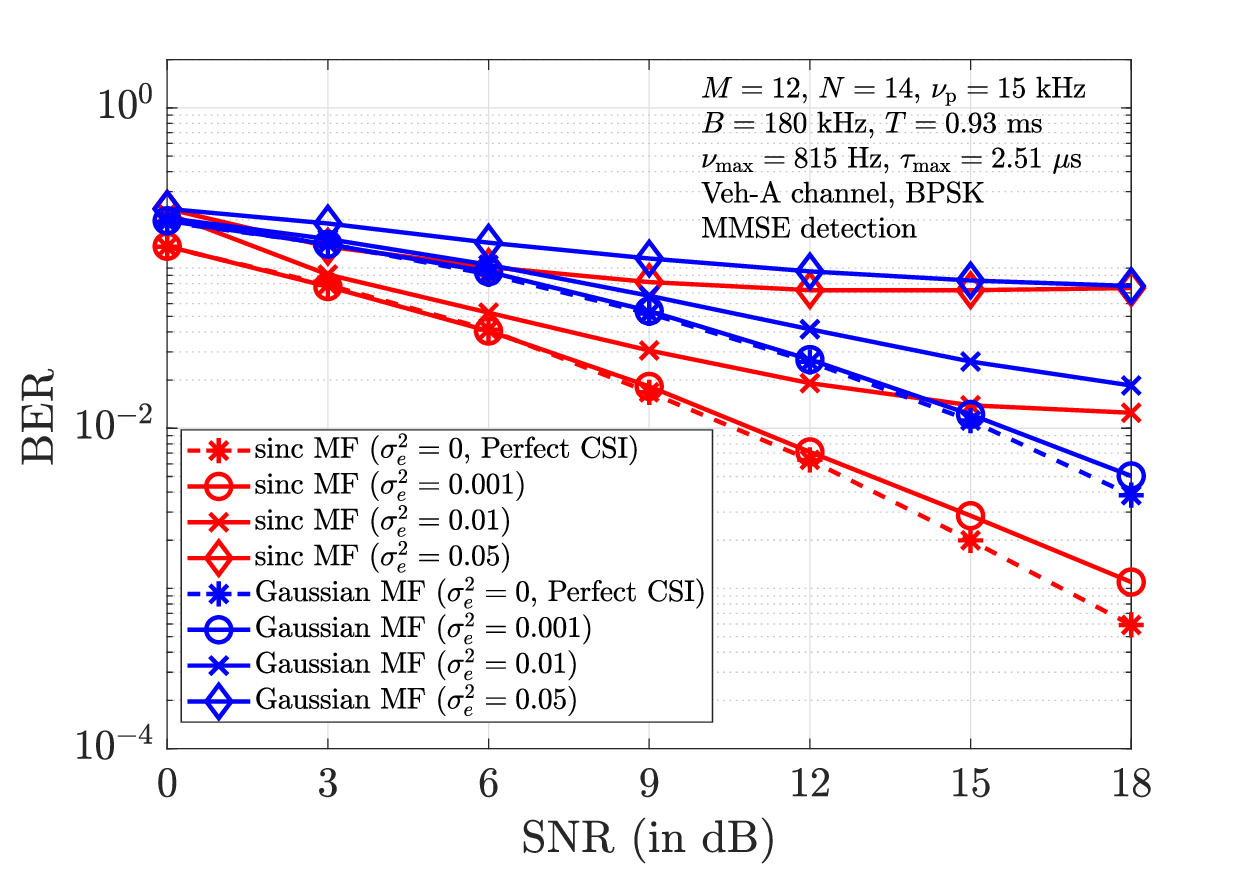}
\caption{BER performance comparison between sinc and Gaussian filters in Zak-OTFS with MMSE detection for $M=12$, $N=14$, $P=6$, under imperfect CSI.}
\label{fig:impCSI_12x14_mmse}
\vspace{-5mm}
\end{figure}

In Fig. \ref{fig:impCSI_12x14_mmse}, we assess the effect of imperfect channel state information (CSI) on the BER performance. Towards this, we consider the estimation error model $\hat{h}_i = h_i + e_i$, where $e_i$ is the error in the estimate of the channel gain, modeled as $e_i \sim \mathcal{CN}(0, \sigma_e^2)$. BER plots for different estimation error variances ($\sigma_e^2=0.001,0.01,0.05$) are shown for sinc and Gaussian filters with MMSE detection. BER plot with perfect CSI is also shown. The results indicate that the sinc filter outperforms the Gaussian filter for the considered error variances, which is consistent with perfect CSI results. Furthermore, as $\sigma_e^2$ increases, both filters exhibit performance degradation, with sinc filter showing a sharper performance decline compared to the Gaussian filter, suggesting its higher sensitivity to estimation inaccuracies.

The derived closed-form expressions for the noise covariance for different filters/configurations can be used to whiten the noise and obtain the performance of maximum-likelihood (ML) detection as follows.
The received signal vector ${\bf y}$ is given by (\ref{sys_mod}), i.e., $\bf y = \bf Hx + \bf n$, where $\bf n$ is the correlated noise vector with the covariance matrix ${\bf C}_{\mathrm n}$. The noise-whitened received vector, denoted by $\tilde{\bf y}$, is given by
\begin{equation}
\tilde{\bf y} \hspace{0.5mm} = \hspace{0.5mm} {\bf R}_{\mathrm n}^{-1}{\bf y} \hspace{0.5mm}  = \hspace{0.5mm} {\bf R}_{\mathrm n}^{-1}{\bf Hx}+{\bf R}_{\mathrm n}^{-1}{\bf n},
\label{eqn:white_sys_model}
\end{equation}
where ${\bf R}_{\mathrm n}\in \mathbb{C}^{MN \times MN}$ is a lower-triangular matrix obtained from Cholesky decomposition of the noise covariance matrix ${\bf C}_{\mathrm n}$, i.e., ${\bf R}_{\mathrm n}{\bf R}_{\mathrm n}^{H}={\bf C}_{\mathrm n}$. The $\bf{R}_{\mathrm n}^{-1}\bf n$ in (\ref{eqn:white_sys_model}) is the whitened noise vector with covariance matrix $\bf I$. Now, the ML detection rule is given by
\begin{equation}
\hat{\bf x}_{\mathrm {ML}}=\underset{{\bf x}\in
{\mathbb A}^{MN}} {\mathrm{argmin}}\ ||\Tilde{\bf y}-{\bf R}_{\mathrm n}^{-1}{\bf Hx}||_{2}.
\label{eqn:ml_det}
\end{equation}
\begin{figure}
\centering
\includegraphics[width=9.5cm,height=6.5cm]{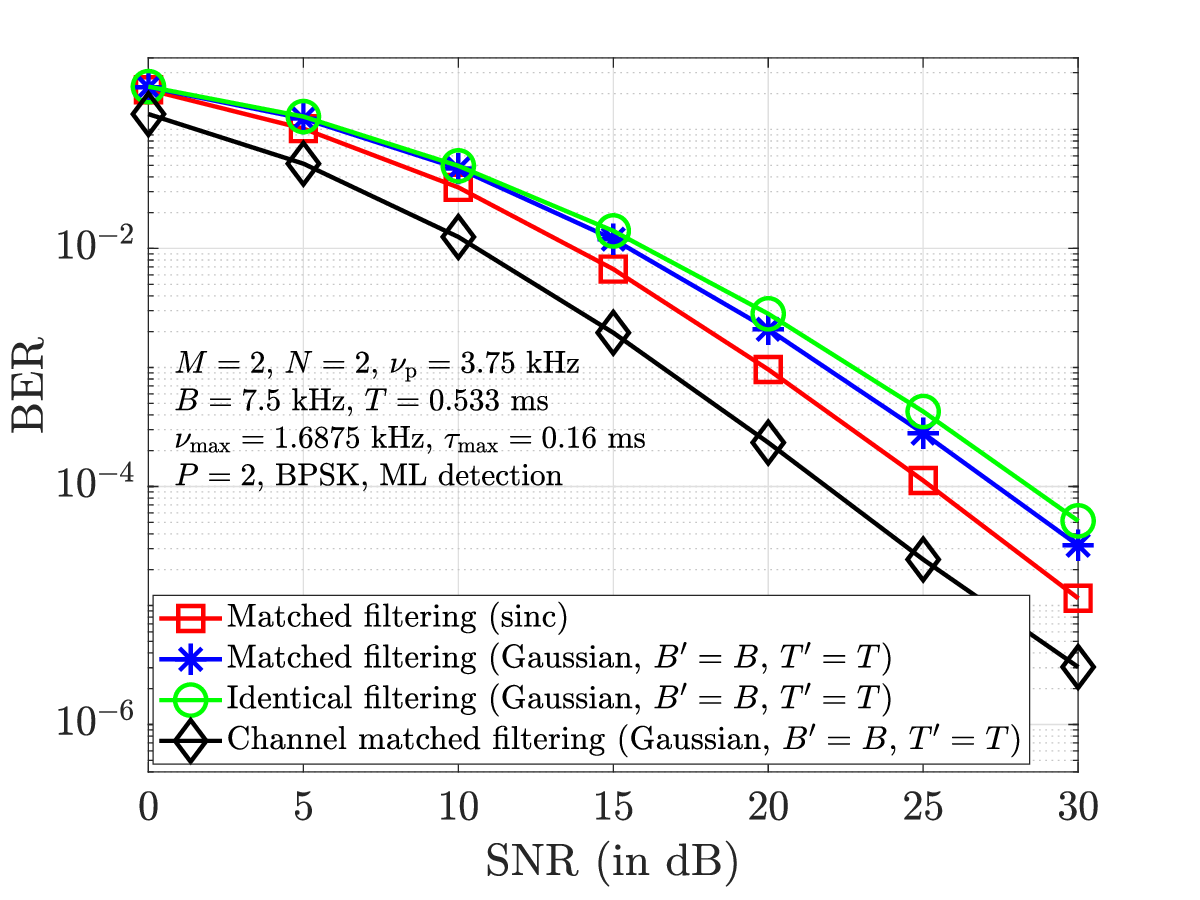}
\vspace{-6mm}
\caption{BER performance comparison between different filters/configurations in Zak-OTFS  
under ML detection for $M=N=2$, $P=2$, and BPSK.}
\label{fig:min_dis}
\vspace{-6mm}
\end{figure}
Since the ML detection complexity in (\ref{eqn:ml_det}) is exponential in the frame size (which is prohibitively complex for large $M$ and $N$), we evaluated the ML BER performance for a system with a small frame size.  Figure \ref{fig:min_dis} shows the simulated ML BER performance of sinc and Gaussian filters in Zak-OTFS with matched filtering for $M=N=2$, $\nu_{\mathrm p} = 3.75$ kHz, BPSK, and $P=2$ paths with a fractional DD profile given by $\{(\tau_1=0.6\frac{\tau_{\mathrm{p}}}{M},\nu_1=0.7\frac{\nu_{\mathrm{p}}}{N}),(\tau_2=1.2\frac{\tau_{\mathrm{p}}}{M},\nu_2=0.9\frac{\nu_{\mathrm{p}}}{N})\}$ and uniform PDP. The BER performance of identical filtering and channel matched filtering with Gaussian filter are also plotted. As observed in large frame sizes with MMSE detection, in Fig. \ref{fig:min_dis} also we observe that $i$) with matched filtering, sinc filter achieves better performance compared to Gaussian filter, and $ii$) with Gaussian filter, matched filtering performs slightly better than identical filtering (by about 1 dB at $10^{-4}$ BER), and channel matched filter performs better than both identical and matched filtering (by about 5 to 6 dB at $10^{-4}$ BER). 

\begin{figure*}
\centering
\subfloat[Sinc, point pilot frame]  {\includegraphics[width=4.35cm,height=4.0cm]{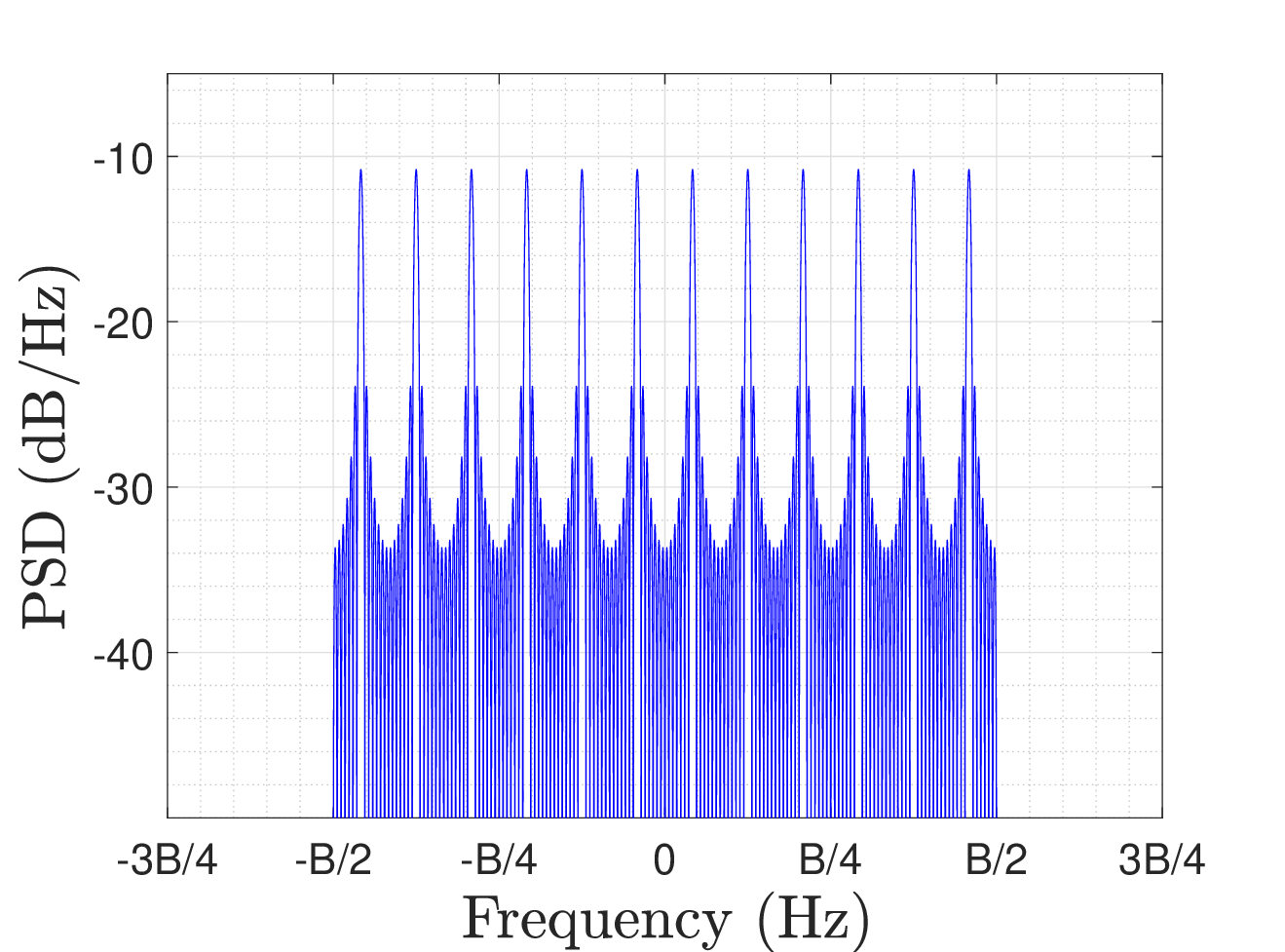} 
\label{fig:motiv_a}
} 
\subfloat[Sinc, data frame]
{\includegraphics[width=4.35cm, height=4.0cm]{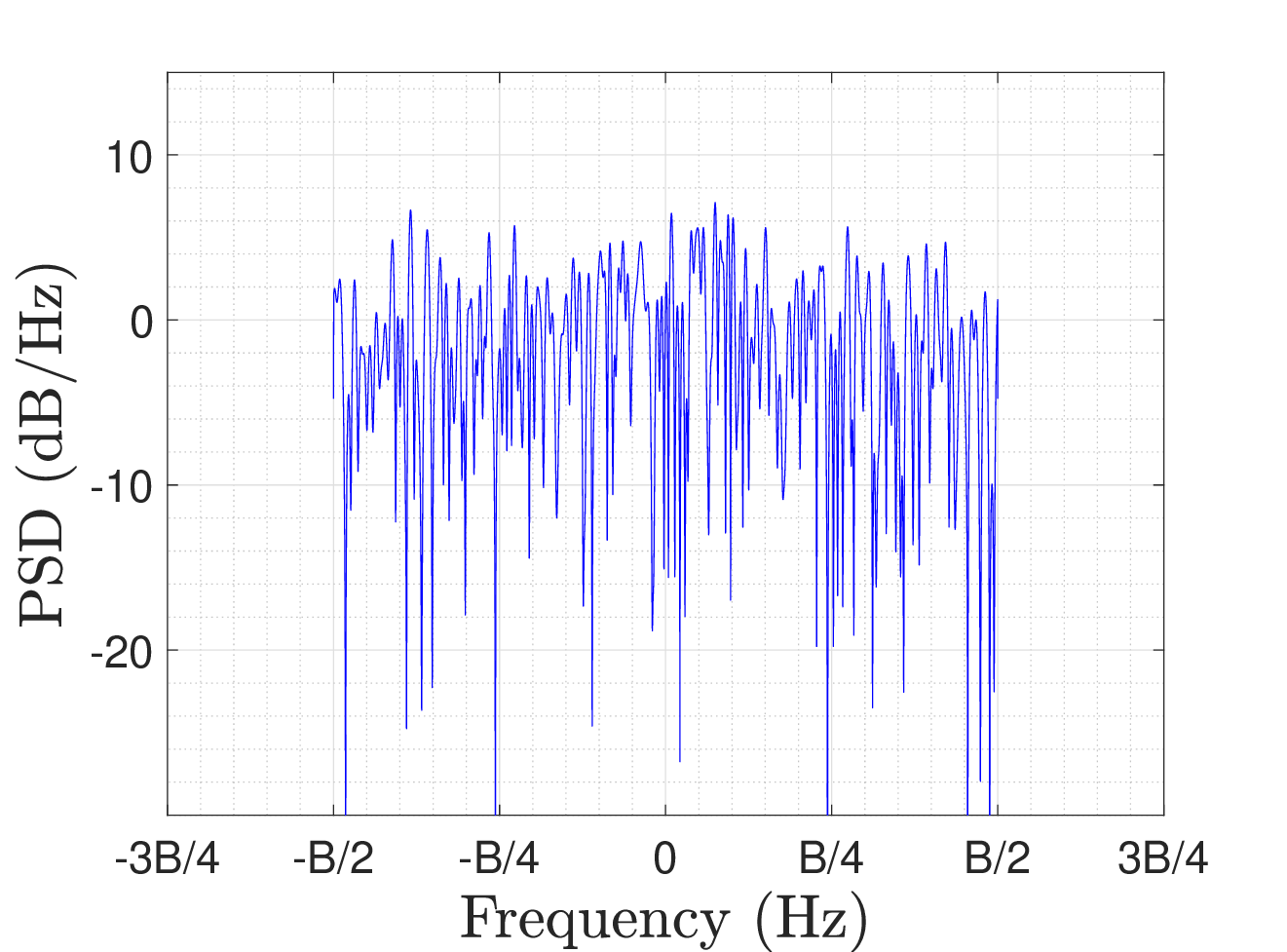} 
\label{fig:motiv_b}
}
\subfloat[Gaussian, point pilot frame]
{\includegraphics[width=4.35cm, height=4.0cm]{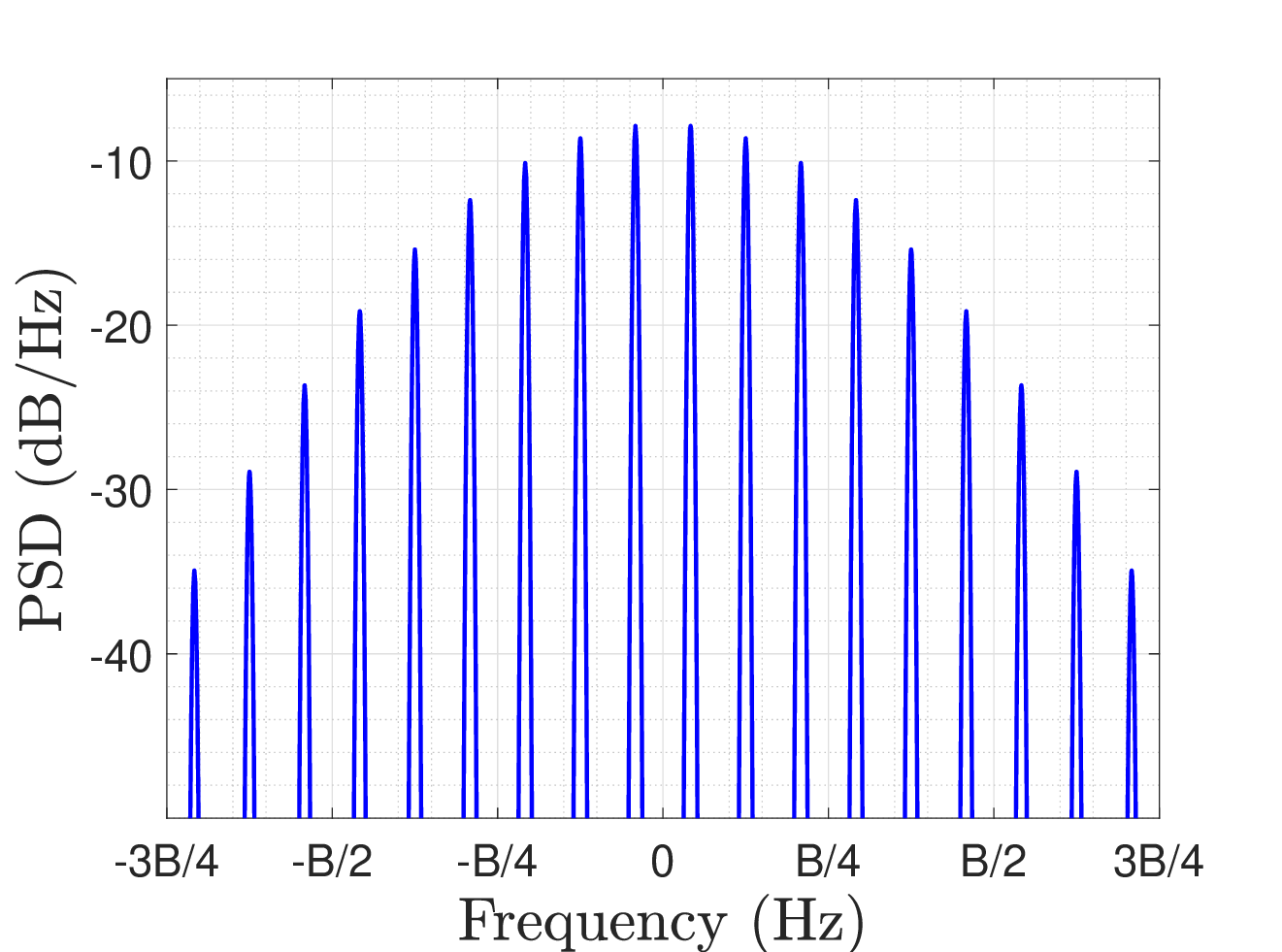} 
\label{fig:motiv_c}
}
\subfloat[Gaussian, data frame]
{\includegraphics[width=4.35cm, height=4.0cm]{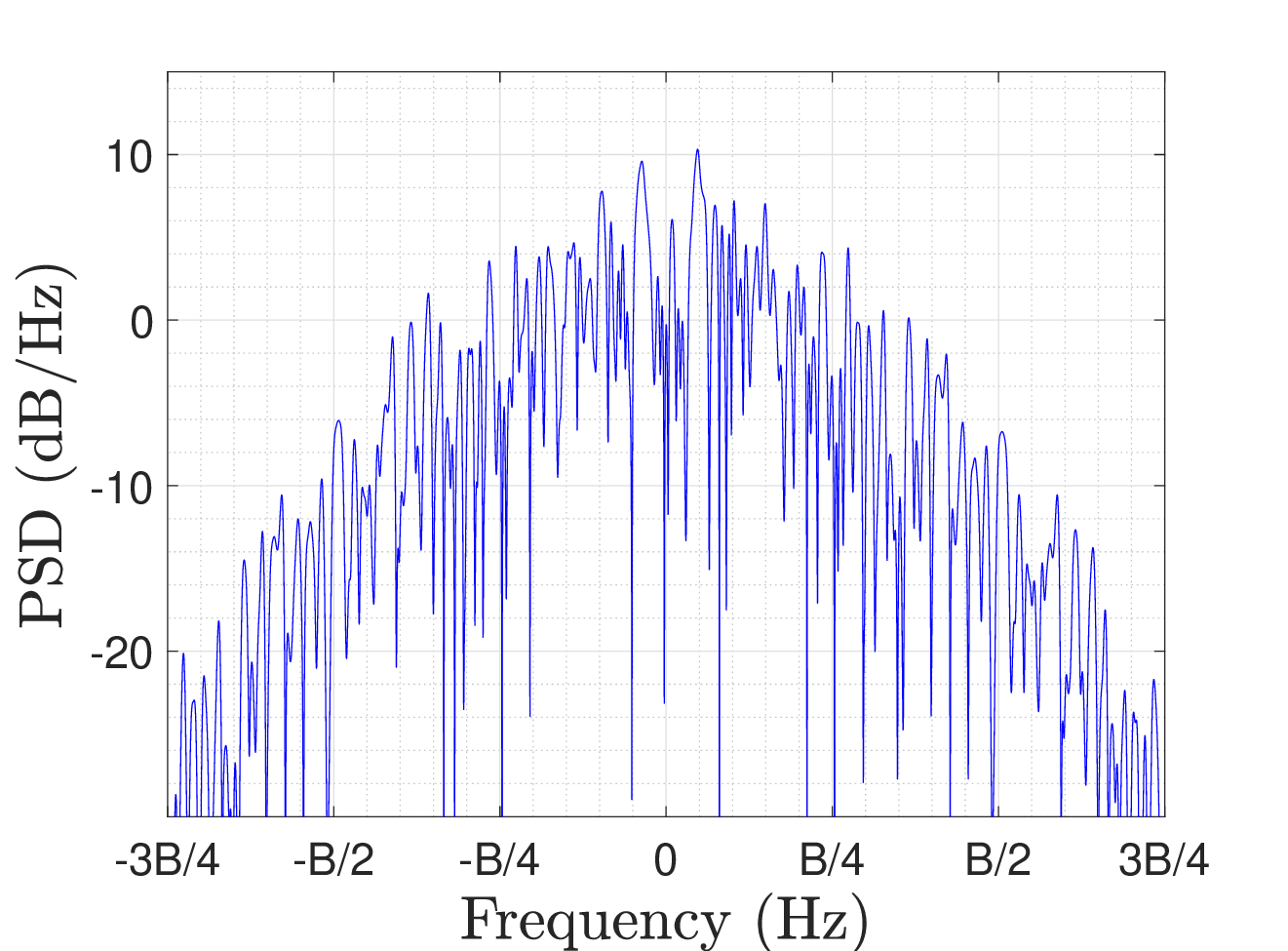} 
\label{fig:motiv_d}
}
\caption{PSD plots of the DD domain signal after Tx filter for (a) sinc filter, point pilot frame, (b) sinc filter, data frame, (c) Gaussian filter, point pilot frame, and (d) Gaussian filter, data frame.} 
\label{fig:motiv}
\vspace{-6mm}
\end{figure*}

\vspace{-2mm}
\subsection{Out-of-band emission characteristics}
In this subsection, we present the out-of-band emission characteristics of the sinc and Gaussian filters. For this, we need to obtain the frequency domain representation of the DD domain signal at the output of the Tx filter. $MN$ symbols,
$x[k,l]$, $0\leq k\leq M-1, 0\leq l\leq N-1$, are placed inside the fundamental period $\mathcal{D}_{0}$. Owing to the quasi-periodicity, the DD domain representation of the $(k,l)$th symbol becomes
\begin{eqnarray}
\hspace{-5mm}
x_{\mathrm{dd},k,l}(\tau,\nu) & \hspace{-2mm} = & \hspace{-2mm} x[k,l]\sum_{n,m\in \mathbb{Z}}\delta\Big(\tau-\frac{k\tau_{\mathrm{p}}}{M}-n\tau_{\mathrm{p}}\Big) \nonumber \\
& \hspace{-0mm} & \hspace{15mm}
\delta\Big(\nu-\frac{l\nu_{\mathrm{p}}}{N}-m\nu_{\mathrm{p}}\Big)e^{j2\pi\frac{nl}{N}}.
\end{eqnarray}
The entire DD frame is given by $x_{\mathrm{dd}}(\tau,\nu)=\sum_{k=0}^{M-1}\sum_{l=0}^{N-1}x_{\mathrm{dd},k,l}(\tau,\nu)$. After passing through the Tx filter $w_{\mathrm{tx}}(\tau,\nu)=w_1(\tau)w_2(\nu)$, the DD domain signal becomes
\begin{eqnarray}
\hspace{-5mm}
x_{\mathrm{dd}}^{w_{\mathrm{tx}}}(\tau,\nu) & \hspace{-2mm} = & \hspace{-2mm}w_{\mathrm{tx}}(\tau,\nu)*_{\sigma}x_{\mathrm{dd}}(\tau,\nu) \nonumber \\
& \hspace{-34mm} = & \hspace{-19mm} \sum_{k=0}^{M-1}\sum_{l=0}^{N-1} \hspace{-1mm} x[k,l]
\hspace{-1mm}
\sum_{n,m\in \mathbb{Z}}
\hspace{-1.5mm}
w_1\Big(\tau-\frac{k\tau_{\mathrm{p}}}{M}-n\tau_{\mathrm{p}}\Big) 
\nonumber \\
& \hspace{-22mm} & \hspace{-22mm}
w_2\Big(\nu \hspace{-0.5mm} - \hspace{-0.5mm} \frac{l\nu_{\mathrm{p}}}{N} \hspace{-0.5mm} - \hspace{-0.5mm} m\nu_{\mathrm{p}}\Big)
e^{j2\pi\nu\left(\frac{k\tau_{\mathrm{p}}}{M}+n\tau_{\mathrm{p}}\right)}e^{-j2\pi\frac{k\tau_{\mathrm{p}}}{M}\left(\frac{l\nu_{\mathrm{p}}}{N}+m\nu_{\mathrm{p}}\right)}\hspace{-0.5mm}.
\end{eqnarray}
The frequency domain representation of a DD domain signal can be obtained via inverse frequency-Zak transform\footnote{Inverse frequency-Zak transform of a DD function $a(\tau,\nu)$ is defined as $Z_{f}^{-1}(a(\tau,\nu))=\sqrt{\nu_{\mathrm{p}}}\int_{0}^{\tau_{\mathrm{p}}}a(\tau,f)e^{-j2\pi f\tau}d\tau.$}. Therefore, the frequency domain Tx filtered signal is given by
\begin{eqnarray}
X(f) & \hspace{-2mm} = & \hspace{-2mm} Z_{f}^{-1}\big(x_{\mathrm{dd}}^{w_{\mathrm{tx}}(\tau,\nu)}\big) \nonumber \\
& \hspace{-13mm} = & \hspace{-8mm}\sqrt{\nu_{\mathrm{p}}}\sum_{k=0}^{M-1}\sum_{l=0}^{N-1} x[k,l]\sum_{m\in \mathbb{Z}}w_2\Big(f-\frac{l\nu_{\mathrm{p}}}{N}-m\nu_{\mathrm{p}}\Big) \nonumber \\
& \hspace{-13mm} & \hspace{-8mm} e^{-j2\pi\frac{k\tau_{\mathrm{p}}}{M}\big(\frac{l\nu_{\mathrm{p}}}{N}+m\nu_{\mathrm{p}}\big)} \nonumber \\
& \hspace{-13mm} & \hspace{-8mm}
\underbrace{\sum_{n\in\mathbb{Z}}\int_{0}^{\tau_{\mathrm{p}}}w_1\Big(\tau-\frac{k\tau_{\mathrm{p}}}{M}-n\tau_{\mathrm{p}}\Big)e^{-j2\pi f\big(\tau-\frac{k\tau_{\mathrm{p}}}{M}-n\tau_{\mathrm{p}}\big)}d\tau}_{\overset{\Delta}{=}W_1(f)} \nonumber \\
& \hspace{-13mm} = &  \hspace{-8mm} \sqrt{\nu_{\mathrm{p}}}\hspace{1mm}W_1(f)\sum_{k=0}^{M-1}\sum_{l=0}^{N-1} x[k,l]\sum_{m\in \mathbb{Z}}w_2\Big(f-\frac{l\nu_{\mathrm{p}}}{N}-m\nu_{\mathrm{p}}\Big) \nonumber \\
& \hspace{-13mm} & \hspace{-8mm}
e^{-j2\pi\frac{k\tau_{\mathrm{p}}}{M}\big(\frac{l\nu_{\mathrm{p}}}{N}+m\nu_{\mathrm{p}}\big)},
\label{eqn:psd}
\end{eqnarray}
where $W_1(f)$ is the frequency domain representation of the delay domain component of the Tx filter $w_1(\tau)$. Using (\ref{eqn:psd}), we have computed the PSD of the Tx filter output signal for sinc and Gaussian filters. Figure \ref{fig:motiv} shows these PSD plots for two types of frames: 1) a point pilot frame where $x[k,l]=1$ for $k=M/2$, $l=N/2$ and zero otherwise, and 2) a data frame where $x[k,l]$, $k=0,\cdots,M-1$, $l=0,\cdots,N-1$ are drawn from BPSK alphabet. Figures \ref{fig:motiv}(a) and (b) show the PSD plots for point pilot frame and data frame, respectively, for sinc filter. Likewise, Figs. \ref{fig:motiv}(c) and (d) show the PSD plots for point pilot frame and data frame, respectively, for Gaussian filter. It can be observed that the PSD for sinc filter is fully contained within the range $(-B/2, B/2)$. Whereas, the Gaussian filter has components outside of the $(-B/2,B/2)$ range. For the chosen value of $\alpha_{\tau}=1.584$ for Gaussian filter, the leakage outside $(-B/2,B/2)$ is close to $1 \%$.

\section{Conclusion}
\label{sec5}
We derived discrete DD domain closed-form expressions for the end-to-end I/O relation and noise covariance in Zak-OTFS modulation for different DD domain filters employed at the transmitter and receiver. We considered sinc and Gaussian DD filters at the transmitter, and identical filtering (Rx filter same as Tx filter), matched filtering (Rx filter matched to Tx filter), and channel matched filtering (Rx filter matched to the cascade of Tx filter and channel) at the receiver. Except for the case of identical filtering with sinc filter, we derived exact closed-form expressions. For identical filtering with sinc filter, we obtained approximate closed-form expressions which are shown to be accurate. For easy and immediate reference, we have summarized the equation numbers of the closed-form expressions in Table \ref{tab_summary}. 
The derived closed-form expressions eliminate the need to numerically compute the multiple integrals involved in the cascade of twisted convolution operations in the Zak-OTFS transceiver, which, in turn, significantly reduced system simulation run times. Our simulation results showed that, while matched filtering achieved slightly better or almost the same performance as identical filtering, channel matched filtering achieved the best performance among the three. In this work, we have considered sinc and Gaussian filters which are commonly considered in the Zak-OTFS literature (e.g., \cite{ref4},\cite{pulse_shaping}). The derivation of closed-form expressions for RRC filter remains open. Derivation of closed-form expressions for other types of filters, including the RRC filter, can be taken up for future work. Also, transceiver algorithms for Zak-OTFS detection/channel estimation, and design of efficient pilot schemes in both single-user and multiuser scenarios can be taken up for future investigation.
\vspace{0mm}
\begin{table}[t]
\centering
\vspace{5mm}
\begin{tabular}{|l||c|c||c|c|}
\hline
  &  \multicolumn{4}{|c|}{Eq. nos. of closed-form expressions} \\ \cline{2-5}
Tx/Rx filtering & \multicolumn{2}{c||}{sinc filter} & \multicolumn{2}{c|}{Gaussian filter} \\ \cline{2-5}
& \multicolumn{1}{c|}{$h_{\mathrm{eff}}[k,l]$} & \multicolumn{1}{c||}{Noise} & \multicolumn{1}{c|}{$h_{\mathrm{eff}}[k,l]$} & \multicolumn{1}{c|}{Noise} \\ 
& \multicolumn{1}{c|}{} & \multicolumn{1}{c||}{covariance} & \multicolumn{1}{c|}{} & \multicolumn{1}{c|}{covariance} \\ \hline \hline
Identical            & 
(\ref{eqn:sinc_identical_closed})
&  (\ref{eqn:noise_approx_closed}) & 
(\ref{eqn:gauss_identical_discrete}) &(\ref{eqn:gaussian_noise_covariance})   \\ \hline
Matched  & (\ref{eqn:MF_IO_sinc})     & (\ref{eqn:MF_covar_sinc})  & (\ref{eqn:MF_IO_Gauss})  & (\ref{eqn:MF_covar_Gauss}) \\ \hline
Channel matched  & (\ref{eqn:CMF_IO_sinc}) & (\ref{eqn:CMF_covar_sinc}) & (\ref{eqn:gauss_eff_channel_matched_samples}) & (\ref{eqn:CMF_covar_Gauss})      \\ \hline
\end{tabular}
\caption{Summary of Eq. nos. of closed-form expressions.} 
\vspace{-6mm}
\label{tab_summary}
\end{table}

\appendices
\section{Derivation of (\ref{eqn:sinc_identical_closed})}
\label{appendix_a1}
The term $(T-|x|)\mathrm{sinc}((T-|x|)(\nu-\nu_{i}))$ inside the integral of $(\ref{eqn:Exact_identical_channel})$ 
is expanded as
\begin{eqnarray}
(T-|x|)\mathrm{sinc}((T-|x|)\underbrace{(\nu-\nu_{i})}_{\overset{\Delta}{=} \nu'_i}) & \hspace{-2mm} = & \hspace{-2mm} T\mathrm{sinc}(T\nu'_i)\underbrace{\mathrm{cos}(\pi|x|\nu'_i)}_{=\mathrm{cos}(\pi x\nu'_i)}\nonumber \\
& \hspace{-18mm} & \hspace{-18mm}  - \hspace{0.5mm} \mathrm{cos}(\pi T\nu'_i)|x|\mathrm{sinc}(\pi|x|\nu'_i).
\label{eqn:closed_expansion}
\end{eqnarray}
Substituting (\ref{eqn:closed_expansion}) in
(\ref{eqn:Exact_identical_channel}) gives the integral in (\ref{eqn:Exact_identical_channel}) as
\begin{eqnarray}
I_i(\tau,\nu)  & \hspace{-2mm} \overset{\Delta}{=} & \hspace{-2mm} \int_{-T}^{T}e^{-j\pi x(\nu+\nu_{i})}\mathrm{sinc}(B(x+\tau)) \nonumber \\
& \hspace{-2mm} & \hspace{-2mm} \mathrm{sinc}(B(x+\tau_{i})) (T-|x|)\mathrm{sinc}((T-|x|)\nu_{i}')dx \nonumber \\
& \hspace{-2mm} = & \hspace{-2mm} T\mathrm{sinc}(T\nu_{i}')\bigg(\int_{-T}^{T}\alpha_{i}(x)dx\bigg) \nonumber \\
& \hspace{-2mm} & \hspace{-2mm} -\hspace{0.5mm} \mathrm{cos}(\pi T\nu_i')\bigg(\int_{-T}^{T}\beta_{i}(x)dx\bigg),
\label{eqn:sinc_ident_integral_1}
\end{eqnarray}
where $\alpha_i(x)$ and $\beta_i(x)$ are given by
\begin{eqnarray}
\alpha_i(x) & \hspace{-2mm} = & \hspace{-2mm} e^{-j\pi x(\nu+\nu_{i})}\mathrm{sinc}(B(x+\tau))\mathrm{sinc}(B(x+\tau_{i}))\nonumber \\
& \hspace{-2mm} & \hspace{-2mm} \mathrm{cos}(\pi x\nu'_i),
\end{eqnarray}
\begin{eqnarray}
\beta_i(x) & \hspace{-2mm} = & \hspace{-2mm} e^{-j\pi x(\nu+\nu_i)}\mathrm{sinc}(B(x+\tau))\mathrm{sinc}(B(x+\tau_{i})) \nonumber \\
& \hspace{-2mm} & \hspace{-2mm} |x|\mathrm{sinc}(\pi|x|\nu_{i}').
\end{eqnarray}
It is observed that the contribution of the integral of $\beta_i(x)$ in (\ref{eqn:sinc_ident_integral_1}) is much small compared to that of the integral of $\alpha_i(x)$ for
large $M$ and $N$ (consequently, large $B$ and $T$) and operation in the crystalline regime  (i.e., $\tau_{\mathrm{max}} << \tau_{\mathrm p}$ and $2\nu_{\mathrm{max}} << \nu_{\mathrm p}$ \cite{ref3}). 
 Hence, (\ref{eqn:sinc_ident_integral_1}) can be approximated as 
\begin{eqnarray}
I_i(\tau,\nu) & \hspace{-2mm} \approx & \hspace{-2mm} T\mathrm{sinc}(T\nu'_i)\left(\int_{-T}^{T}\alpha_i(x)dx\right) \nonumber \\
& = & T\mathrm{sinc(T\nu'_i)}\int_{-T}^{T}e^{-j\pi x(\nu+\nu_{i})}\mathrm{sinc}(B(x+\tau)) \nonumber \\
& & \mathrm{sinc}(B(x+\tau_{i})) 
\mathrm{cos}(\pi x\nu'_i)dx \nonumber \\
& \hspace{-15mm} \approx & \hspace{-10mm} T\mathrm{sinc(T\nu'_i)}\int_{-\infty}^{\infty}e^{-j\pi x(\nu+\nu_{i})} \nonumber \\
& \hspace{-15mm} & \hspace{-10mm} \underbrace{\mathrm{sinc}(B(x+\tau))\mathrm{sinc}(B(x+\tau_{i}))}_{\overset{\Delta}{=}p_{i}(x)}\underbrace{\mathrm{cos}(\pi x\nu_{i}')}_{\overset{\Delta}{=}q_{i}(x)}dx.
\label{eqn:sinc_ident_integral_2}
\end{eqnarray}
For large $M$ and $N$, the effect of changing the limit of integration from $[-T, T]$ to $(-\infty, \infty)$ in (\ref{eqn:sinc_ident_integral_2}) is negligible, since the function $p_{i}(x)$ is comparatively very small outside the range $[-T, T]$. So, (\ref{eqn:sinc_ident_integral_2}) can be written as
\begin{eqnarray}
I_i(\tau,\nu) & \hspace{-2mm} \approx & \hspace{-2mm} T\mathrm{sinc}(T\nu_{i}')\int_{-\infty}^{\infty}p_{i}(x)q_{i}(x)e^{-j\pi(\nu+\nu_{i})x}dx \nonumber \\
& \hspace{-25mm} = & \hspace{-14mm} T\mathrm{sinc}(T\nu_{i}')(P_{i}(f)*Q_{i}(f))|_{f=\frac{\nu+\nu_{i}}{2}} \nonumber \\
& \hspace{-25mm} = & \hspace{-14mm} T\mathrm{sinc}(T\nu_{i}')\bigg[P_{i}(f)*\frac{1}{2}\bigg(\delta\bigg(f-\frac{\nu'_i}{2}\bigg) \nonumber \\
& \hspace{-25mm}  & \hspace{-14mm}
+\delta\bigg(f+\frac{\nu'_i}{2}\bigg)\bigg)\bigg]\bigg|_{f=\frac{\nu+\nu_i}{2}} \nonumber \\
& \hspace{-25mm} = & \hspace{-15mm} \left(\frac{T}{2}\right)\mathrm{sinc}(T\nu'_i) \Bigg[P_{i}\bigg(f-\frac{\nu'_i}{2}\bigg)+P_{i}\bigg(f+\frac{\nu'_i}{2}\bigg)\bigg]\bigg|_{f=\frac{\nu+\nu_i}{2}} \nonumber \\
& \hspace{-25mm} = & \hspace{-15mm} \left(\frac{T}{2}\right)\mathrm{sinc}(T\nu'_i)\left(P_{i}(\nu_{i})+P_{i}(\nu)\right),
\label{eqn:sinc_ident_integral}
\end{eqnarray} 
where $P_{i}(f)$ and $Q_{i}(f)$ denote the Fourier transforms of the functions $p_i(x)$ and $q_i(x)$, respectively. $P_i(f)$ is given by 
\begin{eqnarray}
P_{i}(f) & \hspace{-2mm} = & \hspace{-2mm} \int_{-\infty}^{\infty}\mathrm{sinc}(B(x+\tau))\mathrm{sinc}(B(x+\tau_{i}))e^{-j2\pi fx}dx \nonumber \\
& \hspace{-2mm} = & \hspace{-2mm} e^{j\pi f(\tau+\tau_{i})}\Big(\frac{B-|f|}{B^{2}}\Big)\mathrm{sinc}((B-|f|)(\tau-\tau_{i})) \nonumber \\
& \hspace{-2mm} & \hspace{-2mm} \mathbbm{1}_{\{-B<f<B\}}.
\end{eqnarray}
Substituting (\ref{eqn:sinc_ident_integral}) in (\ref{eqn:Exact_identical_channel}) gives
\begin{eqnarray}
h_{\mathrm{eff}}(\tau,\nu) & \hspace{-2mm} \approx & \hspace{-2mm} \left(\frac{B}{2}\right)\sum_{i=1}^{P}h_{i}e^{-j2\pi\tau_i\nu_i}\mathrm{sinc}(T(\nu-\nu_{i})) \nonumber \\
& \hspace{-2mm} & \hspace{-2mm} 
(P_{i}(\nu)+P_{i}(\nu_{i})).
\label{eqn:approx_channel}
\end{eqnarray}
Sampling (\ref{eqn:approx_channel}) on 
$\Lambda_{\mathrm{dd}}=\{(k\frac{\tau_{p}}{M},l\frac{\nu_{p}}{N}) | k,l\in \mathbb{Z}\}$ 
gives (\ref{eqn:sinc_identical_closed}) and (\ref{eqn:P_ki_discrete}).

\section{Derivation of
(\ref{eqn:gauss_identical_discrete})}
\label{appendix_b1}
Making the substitution $x=\tau_{1}-\tau$, the integral in (\ref{eqn:gauss_identical_channel}) can be written as 
\begin{eqnarray}
\int e^{-\alpha_{\tau}B^{2}(x+\tau)^{2}}e^{-\alpha_{\tau}B^{2}(x+\tau_{i})^{2}}e^{-j2\pi\nu_{i}(x+\tau)} & & \nonumber    \\ 
e^{-\frac{\alpha_{\nu}T^{2}}{2}\left((\nu-\nu_{i})^{2}+j2\pi\frac{ (\nu-\nu_{i})x}{\alpha_{\nu}T^{2}}+\left(\frac{\pi x}{\alpha_{\nu}T^{2}}\right)^{2}\right)}dx & & \nonumber \\ 
& \hspace{-130mm} = & \hspace{-66mm} \underbrace{\Bigg(\int e^{-\left(2\alpha_{\tau}B^{2}+\frac{\pi^{2}}{2\alpha_{\nu}T^{2}}\right)\left(x+\frac{2\alpha_{\tau}B^{2}(\tau+\tau_{i})+j\pi(\nu+\nu_{i})}{2\left(2\alpha_{\tau}B^{2}+\frac{\pi^{2}}{2\alpha_{\nu}T^{2}}\right)}\right)^{2}}dx\Bigg)}_{=\Big(\frac{\pi}{2\alpha_{\tau}B^{2}+\frac{\pi^{2}}{2\alpha_{\nu}T^{2}}}\Big)^{\frac{1}{2}}} \nonumber \\ 
& \hspace{-0mm} & \hspace{-69mm}  \underbrace{e^{\hspace{-0.5mm}-\left(\hspace{-0.5mm}\alpha_{\tau}B^{2}(\tau^{2}+\tau_{i}^{2})+j2\pi\nu_{i}\tau+\frac{\alpha_{\nu}T^{2}}{2}(\nu-\nu_{i})^{2}-\frac{(2\alpha_{\tau}B^{2}(\tau+\tau_{i})+j\pi(\nu+\nu_{i}))^{2}}{4\left(2\alpha_{\tau}B^{2}+\frac{\pi^{2}}{2\alpha_{\nu}T^{2}}\right)}\hspace{-0.5mm}\right)}}_{\overset{\Delta}{=}e^{-f_{i}(\tau,\nu)}} \nonumber \\ 
& \hspace{-130mm} = & \hspace{-66mm} \left(\frac{\pi}{2\alpha_{\tau}B^{2}+\frac{\pi^{2}}{2\alpha_{\nu}T^{2}}}\right)^{\frac{1}{2}}e^{-f_{i}(\tau,\nu)}.
\label{eqn:Gaussian_integral}
\end{eqnarray}
Substituting (\ref{eqn:Gaussian_integral}) in (\ref{eqn:gauss_identical_channel}), we get $h_{\mathrm{eff}}(\tau,\nu)$ in closed-form as
\begin{equation}
h_{\mathrm{eff}}(\tau,\nu)  =  \left(\frac{2\alpha_{\tau}B^{2}}{2\alpha_{\tau}B^{2}+\frac{\pi^{2}}{2\alpha_{\nu}T^{2}}}\right)^{\hspace{-0.5mm}\frac{1}{2}}\sum_{i=1}^{P}h_{i}e^{-g_{i}(\tau,\nu)},
\label{eqn:gaussian_identical_closed}
\end{equation}
where the DD domain function $g_{i}(\tau,\nu)$ is given by
\begin{eqnarray}
g_{i}(\tau,\nu) & \hspace{-2mm} = & \hspace{-2mm} \alpha_{\tau}B^{2}(\tau^{2}+\tau_{i}^{2})+j2\pi\nu_{i}\tau_{i}+\frac{\alpha_{\nu}T^{2}}{2}(\nu-\nu_{i})^{2} \nonumber \\ 
& & -\frac{(2\alpha_{\tau}B^{2}(\tau+\tau_{i})+j\pi(\nu+\nu_{i}))^{2}}{4\left(2\alpha_{\tau}B^{2}+\frac{\pi^{2}}{2\alpha_{\nu}T^{2}}\right)}.
\label{eqn:gi_tau_nu}
\end{eqnarray}
Sampling (\ref{eqn:gaussian_identical_closed}) on  
$\Lambda_{\mathrm{dd}}=\{(k\frac{\tau_{p}}{M},l\frac{\nu_{p}}{N}) | k,l\in \mathbb{Z}\}$ gives 
(\ref{eqn:gauss_identical_discrete}), where $g_{i}[k,l]$ is the sampled version of $g_{i}(\tau,\nu)$ given by (\ref{eqn:gi_discrete}).

\section{Derivation of
(\ref{eqn:gaussian_noise_covariance})}
\label{appendix_b2}
The term $\mathbb{E}_{\{k_{1},k_{2},q_{1},q_{2}\}}$ in (\ref{eqn:gaussian_noise_sampled}) is given by
\begin{eqnarray}
& & \hspace{-10mm} \mathbb{E}_{\{k_{1},k_{2},q_{1},q_{2}\}}=\iint e^{-\alpha_{\tau}B^{2}\tau_{1}^{2}}e^{-\alpha_{\tau}B^{2}\tau_{2}^{2}} \nonumber \\ 
& & \hspace{-10mm} e^{-\frac{\pi^{2}\left(\frac{k_{1}\tau_{p}}{M}-\tau_{1}+q_{1}\tau_{p}\right)^{2}}{\alpha_{\nu}T^{2}}}e^{-\frac{\pi^{2}\left(\frac{k_{2}\tau_{p}}{M}-\tau_{2}+q_{2}\tau_{p}\right)^{2}}{\alpha_{\nu}T^{2}}}  \nonumber \\ 
& & \hspace{-10mm} \underbrace{\mathbb{E}\bigg[n\hspace{-0.5mm}\left(\frac{k_{1}\tau_{p}}{M}-\tau_{1}+q_{1}\tau_{p}\right)n^{*}\hspace{-0.5mm}\left(\frac{k_{1}\tau_{p}}{M}-\tau_{1}+q_{1}\tau_{p}\right)\bigg]}_{=N_{0}\delta\left(\tau_{2}-\tau_{1}-\frac{(k_{2}-k_{1})\tau_{p}}{M}-(q_{2}-q_{1})\tau_{p}\right)}d\tau_{1}d\tau_{2} \nonumber \\ 
& = & \hspace{-0mm} N_{0}\int e^{-\alpha_{\tau}B^{2}\left(\tau_{1}^{2}+\left(\tau_{1}+\frac{(k_{2}-k_{1})\tau_{p}}{M}+(q_{2}-q_{1})\tau_{p}\right)^{2}\right)} \nonumber \\ 
& & e^{-\frac{\pi^{2}\left(\frac{k_{1}\tau_{p}}{M}-\tau_{1}+q_{1}\tau_{p}\right)^{2}}{\alpha_{\nu}T^{2}}}e^{-\frac{\pi^{2}\left(\frac{k_{1}\tau_{p}}{M}-\tau_{1}+q_{1}\tau_{p}\right)^{2}}{\alpha_{\nu}T^{2}}}d\tau_{1} \nonumber \\ 
& = & \sqrt{\frac{\pi}{2\alpha_{\tau}B^{2}+\frac{2\pi^{2}}{\alpha_{\nu}T^{2}}}}e^{-\frac{g[k_{1},k_{2},q_{1},q_{2}]}{\left(2\alpha_{\tau}B^{2}+\frac{2\pi^{2}}{\alpha_{\nu}T^{2}}\right)}},
\label{eqn:expectation_gaussian_noise}
\end{eqnarray}
where the term $g[k_{1},k_{2},q_{1},q_{2}]$ is given by (\ref{eqn:g_gaussian_noise}). Substituting  
(\ref{eqn:expectation_gaussian_noise}) in (\ref{eqn:gaussian_noise_sampled}) gives the expression in (\ref{eqn:gaussian_noise_covariance}).

\vfill

\end{document}